\documentclass{article}

\usepackage{amsfonts,mathtools}
\usepackage{subfigure}
\usepackage{epsfig}
\usepackage{cancel}
\usepackage{multirow}
\usepackage[usenames]{color}
\usepackage{xcolor}
\usepackage{array}
\usepackage{enumerate}
\usepackage{float}

\usepackage{parskip}
\usepackage{amsmath}

\usepackage{cleveref}
	\crefname{equation}{equation}{equations}
	\crefname{figure}{figure}{figures}	
	\crefname{table}{table}{tables}

\begin{document}

\title{Uncertainty quantification in Eulerian-Lagrangian simulations of (point-)particle-laden 
flows with data-driven and empirical forcing models}

\author{Vasileios Fountoulakis$^a$, H.S. Udaykumar$^b$,  Gustaaf B. Jacobs$^a$ \\
\textit{$^a$Department of Aerospace Engineering, San Diego State} \\ \textit{University, San Diego, CA}\\
\textit{$^a$Department of Mechanical and Industrial Engineering,} \\ \textit{The University of Iowa, Iowa City, IA}\\
corresponding author, gjacobs@sdsu.edu
}

\maketitle
\begin{abstract}
An uncertainty quantification framework is developed for Eulerian-Lagrangian 
models of particle-laden flows, where the fluid is modeled  through
a system of partial differential equations in the Eulerian frame and inertial particles
are traced as points in the Lagrangian frame.
The source of uncertainty in such problems is the particle forcing, which
is determined empirically or computationally with high-fidelity methods (data-driven).
The  framework  relies on the averaging of the
deterministic governing equations with the stochastic forcing 
and allows for an estimation of the first and second moment of the
quantities of interest. 
Via comparison with Monte Carlo simulations,
it is demonstrated that the moment equations
accurately predict the uncertainty for problems
whose Eulerian dynamics are either  governed by  the linear advection
equation or the compressible Euler equations.
In areas of singular particle interfaces and shock singularities
significant uncertainty is generated.
An investigation into the
effect of the numerical methods shows that 
low-dissipative higher-order methods
are necessary to capture  numerical singularities 
(shock discontinuities, singular source terms,
particle clustering)  with low diffusion in the propagation of uncertainty.  
\end{abstract}




\section{Introduction}

Particle-laden flows occur in a number of engineering applications and natural environments, including
aerosol and sprays flows in combustion engines and medical devices, dispersion of environmental pollutants, etc.
In some applications the carrier gas flow travels at  high/supersonic speed 
and shock waves interact with the
particles. Some examples are the dynamics in  
scramjet combustors, detonation of explosives, and volcanic eruptions. 

The number of particles in a process-scale environment  
is too large to model with first-principles models. 
Macro-scale models and
simulation techniques are necessary to simulate such problems.
An Eulerian-Lagrangian (EL) approach combined  with point-particle
modeling is commonly used  to make computations more efficient. In these EL models, the carrier flow is solved in the Eulerian frame,
while the volumeless particles (Particle-Source-In-Cell, or PSIC)  are traced
along their Lagrangian path 
\cite{crowe1977particle}.
The interactions between
the particles and the carrier fluid are modeled though momentum exchange via
singular point source terms \cite{jacobs2009high,jacobs2012high}.

These point particle approaches require  a model for the drag force exerted on the
particles. Under the assumptions of very low Reynolds number, incompressible
flow and spherical particles an analytical solution exists: the Stokes drag
law\cite{stokes1851effect}. For high Reynolds numbers several empirical laws
have been developed. These laws correct the Stokes drag law for various
effects: compressibility \cite{boiko2005drag,loth2008compressibility}, slip
coefficients \cite{tedeschi1999motion}, viscosity ratio for droplets
\cite{feng2001drag}, and others. Generally, physical experiments, canonical
theoretical constructs, and phenomenological arguments are employed to obtain
expressions for the drag forces. A drawback of empirical expressions is the
limited range in parameter space. 

In an ongoing effort, we address this limitation through multi-scaling. 
We use resolved mesoscale simulations to derive
drag force models in a range that applies to an arbitrary macro-scale parameter
range \cite{sen2015evaluation,sen2017evaluation}. This is achieved by
construction of data-driven metamodels from a limited amount of data 
for drag laws, heat transfer, particle-fluid correlations etc. over a wide range
of parameters such as the Mach number, Reynolds number, and particle phase number density.

Both empirical models and meta-models have associated uncertainties, arising
from measurement, numerical errors and sparsity of the data points. 
To understand how this model uncertainty
affects the solution on the macro-scale, one must understand how uncertainty propagates
through the non-linear Eulerian-Lagrangian model. This understanding begins by using
a probabilistic perspective and by modeling the drag force as a random
variable. The quantities of interest (QoIs), such as the flow dynamics and particle dispersion and mixing,
are greatly influenced by the epistemic uncertainty in the drag forces. As a
consequence, the QoIs have to be modeled as random variables also.
This yields a classic stochastic formulation that requires techniques
from  the field of uncertainty
quantification (UQ). 



A proper and complete description of the random variables is through the
estimation of the probability density function, since our quantities of
interest are continuous random variables. To this end, deterministic equations
for the evolution of the probability density function of the quantities of
interest for advection-reaction equations were developed
\cite{venturi2013exact}. Cumulative density function evolution equations of the
QoIs for kinematic wave equations have been developed in
\cite{wang2012uncertainty}. However, these methods remain to be extended to
systems involving  singularities, such as shocks propagating through  a
multiphase mixture with singular sources. 

One of the classic approaches quantifying uncertainties is Monte Carlo (MC).
Even though MC is easy to implement, convergence requires many samples that
makes the method computationally inefficient. Other variants of MC methods that
employ more advanced sampling techniques e.g. Latin hypercube sampling
\cite{iman1980small} and multilevel MC \cite{giles2008multilevel} partially
mitigate this difficulty. 

The stochastic finite element method (SFEM) is a different approach that has
been used for UQ tasks \cite{ghanem1991stochastic}. It is based on the
expansion of the QoIs on the stochastic space using
orthogonal polynomials of the random variables (generalized polynomial chaos
(gPC) \cite{xiu2002wiener}). Another variant of gPC, the multi element
generalized polynomial chaos method, was introduced to deal with the
discontinuities in the stochastic space \cite{wan2005adaptive,wan2006multi}.
However, both methods suffer from their intrusive nature (having to modify the
deterministic code and ending up with a high-dimensional coupled system of
equations)  and the curse of dimensionality. 

In order to deal with the intrusive nature of SFEM researchers proposed
stochastic collocation methods \cite{babuvska2007stochastic}. Some of these
methods, i.e. adaptive sparse grid collocation \cite{ma2009adaptive}, are able
to deal with discontinuities in the stochastic space successfully. In addition,
Gaussian processes \cite{rasmussen2004gaussian} have been also employed to
develop a Bayesian UQ framework that is based on surrogates models
\cite{bilionis2012multi}.

In this paper, we develop a model that propagates the uncertainty of the drag
force through a non-linear coupled EL system. The approach is based on the
averaging of the governing equations and allows for an estimation of the first
moments of uncertainty. The main advantage of the proposed
framework is the simplicity, accuracy, and the computational efficacy. 
This efficiency becomes particularly important when a feedback loop is used 
to reduce the uncertainty in the surrogate models based on uncertainty in the QoI,
a natural extension of the current
work.  In that case, iterations between the QoI and
the drag surrogate model are required that use the computational intensive stochastic macro-models
to reduce the uncertainty in the system. 
A downside of of averaging procedures is that the final system of equations
is not closed. We handle the  closure problem  a priori
by using simulated data. In future efforts, a more consistent  a posteriori
closure will be sought.  

We demonstrate that within this UQ framework
only two moments are required to accurately predict the uncertainty for
problems governed by linear advection and 
the compressible Euler equations in Eulerian frame. We also report
on  the effect of numerical approximation, shock discontinuities, 
singular source terms and particle clustering on the
propagation of uncertainty. 

The paper is organized as follows. In Section 2, the deterministic equations
for the Euler equations and the linearized problems are presented. In Section
3, we derive the stochastic equations. In Section 4, we briefly summarize the
numerical methods employed for the solution of the equations. Numerical tests
and discussion of the results are presented in Section 5. Conclusions and
future research directions are drawn in the Section 6. 

\section{Deterministic models}
\label{sec:main}

We adopt the EL model with the carrier phase governed
by the compressible Euler equations in the Eulerian frame.
This system poses several challenges in terms of mathematical modeling and
numerical predictions. Discontinuous solutions exist because of the presence of
the shocks. In addition, the treatment of the particles as mathematical points
leads to the appearance of singular source terms in the conservation of
momentum. 
These characteristics of shock particle-laden flows also present stiff challenges to
a UQ framework. 
To test and verify the UQ framework, we consider simplifications 
in the following aspects: a) using 1D models, b) linearization of the Eulerian equations and c)
smoothing of the singular particle source terms.
In the following we present the coupled equations that describe the dynamics of the fluid flow and the kinematic and dynamic equations that govern the particle motion. 

\subsection{Deterministic Eulerian-Lagrangian Equations}
\label{section2}

The compressible inviscid flows are governed by the Euler equations which in one-dimensional conservative form are given by:
\begin{equation}\label{eq:euler}
\frac{\partial \mathbf{Q}}{\partial t} + \frac{\partial \mathbf{F}}{\partial x} = \mathbf{S}
\end{equation}
where 
\begin{subequations}
\label{euler_details}
\begin{align}
 \mathbf{Q} &= [\ \rho,\rho u, E \ ]^T , \\
 \mathbf{F} &= [\ \rho u,\rho u^2 + P, (E+P)u \ ]^T ,\\
 \mathbf{S} &= [\ 0,S_m,S_e \ ]^T \label{eq:source}
 \end{align}
\end{subequations}
where $\rho$ is the density of the fluid, $u$ is the velocity of the fluid, $E$ is the energy of the fluid and $P$ is the pressure and given by $P = (\gamma - 1) \left( E - \rho u^2/2 \right),  \gamma = 1.4$.
These equations are closed by the equation of state $T = {\gamma P M^2}/\rho$, where $M = U/ \sqrt{\gamma R T}$ is a reference Mach number determined with the reference velocity ($U$) and reference temperature $T$. The source terms in \cref{eq:source} are given by 
\begin{subequations}
\label{eq:sources}
\begin{align}
 S_m &=  \sum_j^{N_P} f_1 \frac{m_p}{\tau_p}(u_{p_j}-u) \delta (x-x_{p_j}) , \\
 S_e &= \sum_j^{N_P} \left( f_1 \frac{m_p}{\tau_p}(u_{p_j}-u)u_{p_j} + f_2 \frac{m_p}{\tau_p} (T_{p_j}-T) \right) \delta (x-x_{p_j}) 
 \end{align}
\end{subequations}
where $T$ is the fluid temperature, $N_p$ is the number of particles, $\tau_p$
is the particle time constant, $m_{p}$ is the particle mass, $u_{p_j}$ is the
particle velocity, $x_{p_j}$ is the particle position, $T_{p_j}$ is the
particle temperature, $f_1$ and $f_2$ are correction factors. The factor $f_2 =
Nu/(3Pr)$ is determined by the Nusselt number ($Nu$) and the Prandtl number
($Pr$). The empirical correction factor $f_1$ is given by the expression
\cite{boiko1997shock} : 
\begin{equation}
\label{eq:correction} 
f_1 = (18 +
0.285Re_p + 3 \sqrt{Re_p} ) \left( 1 + \exp \left[ - \frac{0.43}{M_f^{4.67}},
\right] \right) \end{equation}
where $Re_p$ is the relative particle Reynolds number and $M_f$ is the relative
particle Mach number. This correction factor is again our source of uncertainty
in the system and is treated as a random variable later on. 

The particles are tracked in the Lagrangian frame. Position of the particles is given by 
\begin{equation}\label{eq:particle_position}
\frac{d x_p}{\partial t}  = u_p
\end{equation}
The particle velocity is given by Newton's second law. The drag force is a combination of the corrected Stokes law for high Reynolds and Mach number and the pressure drag \cite{boiko1997shock}:
\begin{equation} \label{eq:particle_velocity1}
\frac{d u_p}{d t}  = f_1 \frac{u - u_p}{\tau_p} - \frac{1}{\rho_p} \frac{\partial P}{\partial x}
\end{equation}
where $\rho_p$ is the density of the particles. Finally, the particle temperature is calculated from the Fourier's law of heat transfer and the first law of thermodynamics: 
\begin{equation} \label{eq:particle_temperature}
\frac{d T_p}{d t}  = f_2 \frac{T - T_p}{\tau_p} 
\end{equation}

\subsection{Advection equation}

For the purpose of verification and testing, we have
considered a linear advection equation for an Eulerian model whose
solutions do not contain the challenging non-linear behavior of the compressible Euler equation.
This allowed us to focus on the coupling between the Eulerian and Lagrangian system.
We performed tests using the linear advection equation in 1D for the Eulerian model
with a particle interaction source term as follows:
\begin{equation}\label{eq:advection}
\frac{\partial u}{\partial t} + \frac{\partial u}{\partial x} = S_m.
\end{equation}
The particles are tracked individually in their Lagrangian frame. The position of the particle ($x_p$) is given  by  \cref{eq:particle_position}. The velocity of the particle is determined by Newton's second law forced by the drag on the particle:
\begin{equation}\label{eq:particle_velocity}
\frac{d u_p}{d t}  = f_1 \frac{u - u_p}{\tau_p}
\end{equation}
These equations form a system of coupled equations with linear advection behavior and are similar to the coupled system of Euler equations with particles. 

\section{Stochastic models}
\label{sec:experiments}

We use a corrected Stokes law for high Reynolds and Mach numbers to calculate
the drag force exerted on the particles. This drag law is an empirical law, 
therefore subject to uncertainty. To
simulate the uncertainty of this calculation we will treat the empirical
correction factor ($f_1$) as a random variable. This uncertainty is propagated
to the fluid and particles QoIs, which are treated as random
variables as well. 

In the following, we derive the equations for the first couple of moments for
the linear advection and the Euler equations based on the Reynolds and Favre
average procedures respectively. 

\subsection{Advection equations}

Using  Reynolds averaging we derive equations for the first two moments of the EL
system based on linear advection equation. The instantaneous value is 
decomposed as $u = \overline{u} + {u'}$, where $\overline{u}$ is the mean field and ${u'}$ is the fluctuation. The stochastic equations derived for the linear advection equation are the following:
\begin{subequations}
\label{stochastic_linear}
\begin{align}
\frac{\partial \overline{u}}{\partial t} + \frac{\partial \overline{u}}{\partial x}  &=  \overline{ S_m },\\
\frac{\partial \sigma_u}{\partial t} + \frac{\partial \sigma_u}{\partial x}  &=  2 \overline{u S_m} - 2\overline{u} \overline{S_m} = 2 \overline{S_m u'}
 \end{align}
\end{subequations}
where $\overline{(\cdot)}$ denotes the Reynolds averaged quantity and $\sigma_{(\cdot)}$ is the variance of the quantity. To obtain an equation for the variance we used the identity: $\sigma_u = \overline{u^2} - \overline{u}^2$. Similarly, we derive the particle stochastic equation as follows:
\begin{subequations}
\label{stochastic_particles}
\begin{align}
\frac{\partial \overline{x_p}}{\partial t}  &=  \overline{ u_p} ,\\
\frac{\partial \sigma_{x_p}}{\partial t}  &=  2\overline{x_p u_p} - 2 \overline{x_p} \ \overline{u_p}, \\
\frac{d \overline{u_p}}{d t}  &= \overline{  f_1 \frac{u - u_p}{\tau_p} }, \\
\frac{\partial \sigma_{u_p}}{\partial t}  &=  2\overline{  f_1 u_p \frac{u - u_p}{\tau_p} }  - 2 \overline{u_p} \overline{f_1 \frac{u - u_p}{\tau_p}}
 \end{align}
\end{subequations}

\subsection{Euler equations}

In compressible flows, Favre averaging is the most common way to obtain moment equations. Favre averaging $u = \tilde{u} +u''$ is a density-weighted variant of the Reynolds averaging where the mean field is $\tilde{u} = \overline{\rho u}/\overline{\rho}$. The mean field equations are the following:
\begin{subequations}
\label{eq:favre}
\begin{align}
\frac{\partial \overline{\rho}}{\partial t} + \frac{\partial \overline{\rho} \tilde{u}}{\partial x}  &= 0 \\
\frac{\partial \overline{\rho} \tilde{u}}{\partial t} + \frac{\partial (\overline{\rho} \tilde{u}^2 + \overline{P})}{\partial x}  &= \overline{S_m} - \frac{\partial \overline{\rho {u''}^2}}{\partial x} \\
\frac{\partial \overline{E} }{\partial t} + \frac{\partial (\overline{E} + \overline{P}) \tilde{u}}{\partial x}  &= \overline{S_e} - \frac{\partial \overline{ (E+P)u''}}{\partial x}
 \end{align}
\end{subequations}
where $\tilde{(\cdot )}$ denotes the Favre-averaged quantities. 

Here, our focus is on the fluid velocity as a QoI. 
A measure of the velocity uncertainty comes
from its second moment, which in our case is $\widetilde{{u''}^2}$. 
This quantity is related to  the turbulence kinetic energy, $k=\widetilde{{u''}^2}/2$, which is governed by the following equation:
\begin{equation} \label{eq:TKE} \frac{\partial k}{d t} + \tilde{u}
\frac{\partial k}{\partial x} =  - 2k\frac{\partial \tilde{u}}{\partial x} -
\frac{1}{2\overline{\rho}} \frac{d}{dx}  \overline{\rho {u''}^3} -
\frac{1}{\overline{\rho}} \overline{\frac{\partial P}{\partial x} u''} +
\frac{1}{\overline{\rho}} \overline{S_m u''} \end{equation}

All the equations above have terms that include the expected value of the
product of two or more quantities. For example, the source term $\overline{S_m}$
when averaged using (\ref{eq:sources}) leads to:
\begin{equation}
\label{eq:source_averaged}
\overline{S_m} = \sum_j^{N_P} 
\frac{m_p}{\tau_p} 
(\overline{f_1}( \overline{u_{p_j}\delta (x-x_{p_j})} - \overline{u \delta (x-x_{p_j})})
+ \overline{f_1^\prime u_{p_j}^\prime \delta (x-x_{p_j})}
+ \overline{f_1^\prime u^\prime \delta (x-x_{p_j})})
\end{equation}
Terms such as the last two terms in the equation above are in general not known a
priori so the system of equations is not closed. 
To test averaged equations it is common to use a posteriori closure.
In this study,
we choose to follow this approach and determine the closure 
terms from data that are gathered for different
realizations of the correction factor $f_1$. In future work, we
aim to use a  multi-scale framework to provide us with a
closed system of equations that can be used for uncertainty quantification
purposes. Here, we focus on the formulation and testing of the averaged governing
system of equations.

While the equations appear fairly straightforward, there are a number of
characteristics of the governing system of equations that pose challenges for
solving them. We remark on them below:

\subsubsection*{Remark 1} The Dirac distribution function in the deterministic term is singular. If there are isolated particles or clusters of particles in the particle-phase where the particle number density suddenly changes, then the deterministic source term can be singular also. To regularize these a number of techniques exist \cite{suarez2014high,suarez2017regularization}. 

\subsubsection*{Remark 2} 
The averaged source term $\overline{S_m}$ is the sum of the deterministic
source terms in equation \cref{eq:sources}. A singular deterministic source
term can make the averaged source term singular or near singular as well. To
accurately capture it sufficient resolution and large number of samples are
required.

\subsubsection*{Remark 3} Shock singularities in the fluid phase create near-singular averaged source terms also. The derivatives on the right hand side in equation \cref{eq:TKE} are near-singular in shock regions where the pressure, velocity and other fluid variables are discontinuous. These need to be regularized. 

\subsubsection*{Remark 4} Two well-known particle-phase dynamics and dispersion patterns include trajectory crossing where two particles have significantly different velocity at the same location and particle clustering, where particles accumulate in a small area. Both these effects lead to singularities in the moments and/or difficulties in determining and closing the moments \cite{schwarzkopf2011multiphase}.

\subsubsection*{Remark 5} The variance equations include directly or indirectly the variation of the realizations ($u'$ or $u''$) of the fluid phase solution. In regions where the deterministic solution is singular, the variance of the stochastic solution is inherently large and singular. This affects the right hand side in equation \cref{eq:TKE} and the source term in equation \cref{stochastic_linear}. 

\section{Numerical Approximations}

To solve the governing EL system in section \ref{section2},
we employ the Particle-Source-In-Cell (PSIC) method \cite{crowe1977particle,jacobs2009high}. 
In this method, 
the Eulerian models are approximated on a static grid, while
the point particles are traced on a Lagrangian frame.
The influence of the point particle on the carrier phase is distributed over the grid through distribution
functions, while the influence of the carrier phase is interpolated
to the particle position.

Initially, we approximate the advection equation \cref{eq:advection} 
and the stochastic linear equations
\cref{stochastic_linear} 
with a simple first-order upwind scheme on a uniform mesh in space
and temporally with the first-order explicit Euler scheme. To verify the
solution of \cref{stochastic_linear}, we performe MC simulations.
From the realizations of the MC simulations we determined the closure 
source terms a priori and substituted them in the
stochastic equation \cref{eq:advection}, which could then be solved.

Despite
the usage of significant resolution, we find that the combination
of the lower-order upwind schemes lead to significant numerical
diffusion near singular sources in the solution of the stochastic equation
and the match with MC is poor.

We subsequently solve the deterministic and stochastic linear advection equation using the
Chebyshev  collocation method \cite{wissink2018shock}  in space.
For the time discretization we used a fourth-order Runge-Kutta method.  
As compared to the first-order schemes this reduced the dissipation significantly and
lead to a much better match between the MC results and the method of moments.

For the solution of the Euler equations
we used the high-order Weighted - Essentially - Non - Oscillatory (WENO-Z) based  PSIC
algorithm that is described in detail in \cite{jacobs2009high}.

Particle tracking requires a three stage algorithm: 1) locate the host cell of
the particle, 2) interpolate the fluid variables to the particle location, and
3) time advance with a time integration method. To determine the influence
of the particles on the static grid that is used to solve the carrier fluid
equation, we must  approximate the delta function ($\delta(x)$). A normalized
distribution function is used to distribute the influence of the particles onto
the carrier fluid. Here we use either a higher-order spline or a Gaussian
distribution function. 

To avoid aliasing errors the order of interpolation of the fluid quantities at
the particle location must match that of the fluid solver
\cite{jacobs2006high}. Hence, we use a fifth-order essential non-oscillatory
(ENO) interpolation that is tailored for flows containing shock discontinuities
\cite{jacobs2009high}. In addition, a fifth-order spline \cite{abe1986high} is
employed to weight the contribution of the particles onto the fluid solver
grid.

\section{Tests}

To verify the moment equations and understand the propagation of the uncertainty in the coupled
EL system we first focus on an EL system whose Eulerian
dynamics are governed by  a linearized advection equation. 
We subsequently test our framework with the full nonlinear EL system.

\subsection{Advection equation with a single particle}

We begin by taking the source term to be  a single smooth 
exponential function to avoid singularity issues as described
in the Remarks above:
\begin{equation}\label{eq:source_exp}
S_m = - \frac{m_p f_1 (u-u_p)}{\tau_p  \sqrt{2 \pi \sigma^2_\delta} } exp\left[-\frac{(x-x_p)^2}{2 \sigma_\delta^2} \right],
\end{equation}
where $\sigma_\delta^2$ controls the spread of the exponential function. We use a single particle to track the peak of this exponential function and the kinematic equations for this particle is given by \cref{eq:particle_position} and \cref{eq:particle_velocity}.

The domain is set up as $x \in [0,0.07]$.
Periodic boundary conditions are imposed on the boundaries of the domain. 
The particle response time and the total mass of the particles are: $\tau_p =
0.01$ and $m_p=0.002$ respectively.

First, we performed computations with the first-order upwind scheme.
The grid spacing is $\Delta x=1e-5$ and the time step is obeying the CFL
condition $\Delta t=0.4dx$. The length scale of the exponential function is set to
one hundred times the grid spacing $\sigma_\delta = 10^{-3}$. We initialize the
particle at $x_p = 0.02$ at rest. Finally, we assume that the correction factor
follows a Gaussian distribution with a twenty percent error $f_1 \sim
\mathcal{N}(1,0.2)$.

\begin{figure}[htbp]
\begin{center} 
\mbox{
\includegraphics[width=0.45\textwidth]{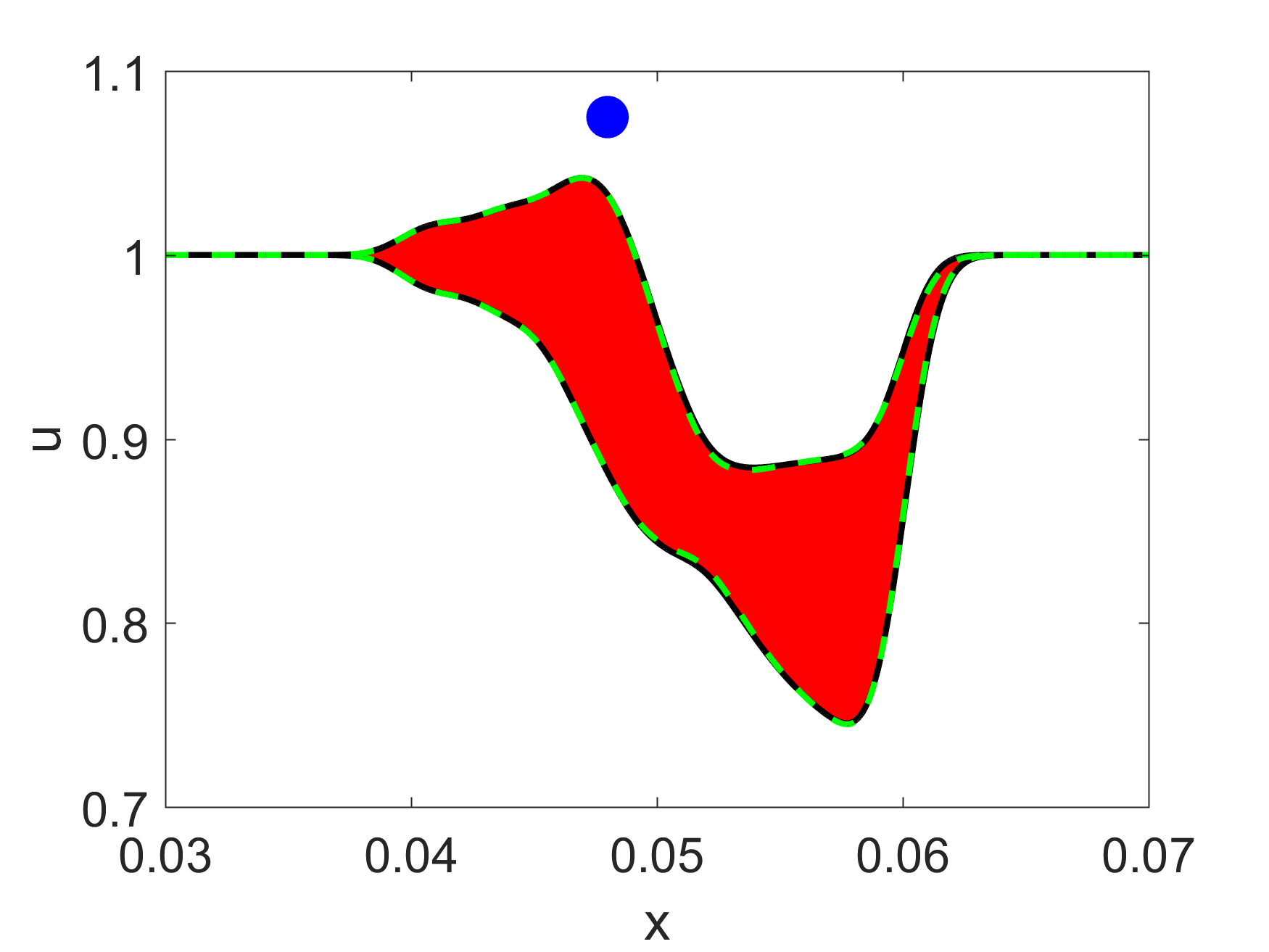}
\includegraphics[width=0.45\textwidth]{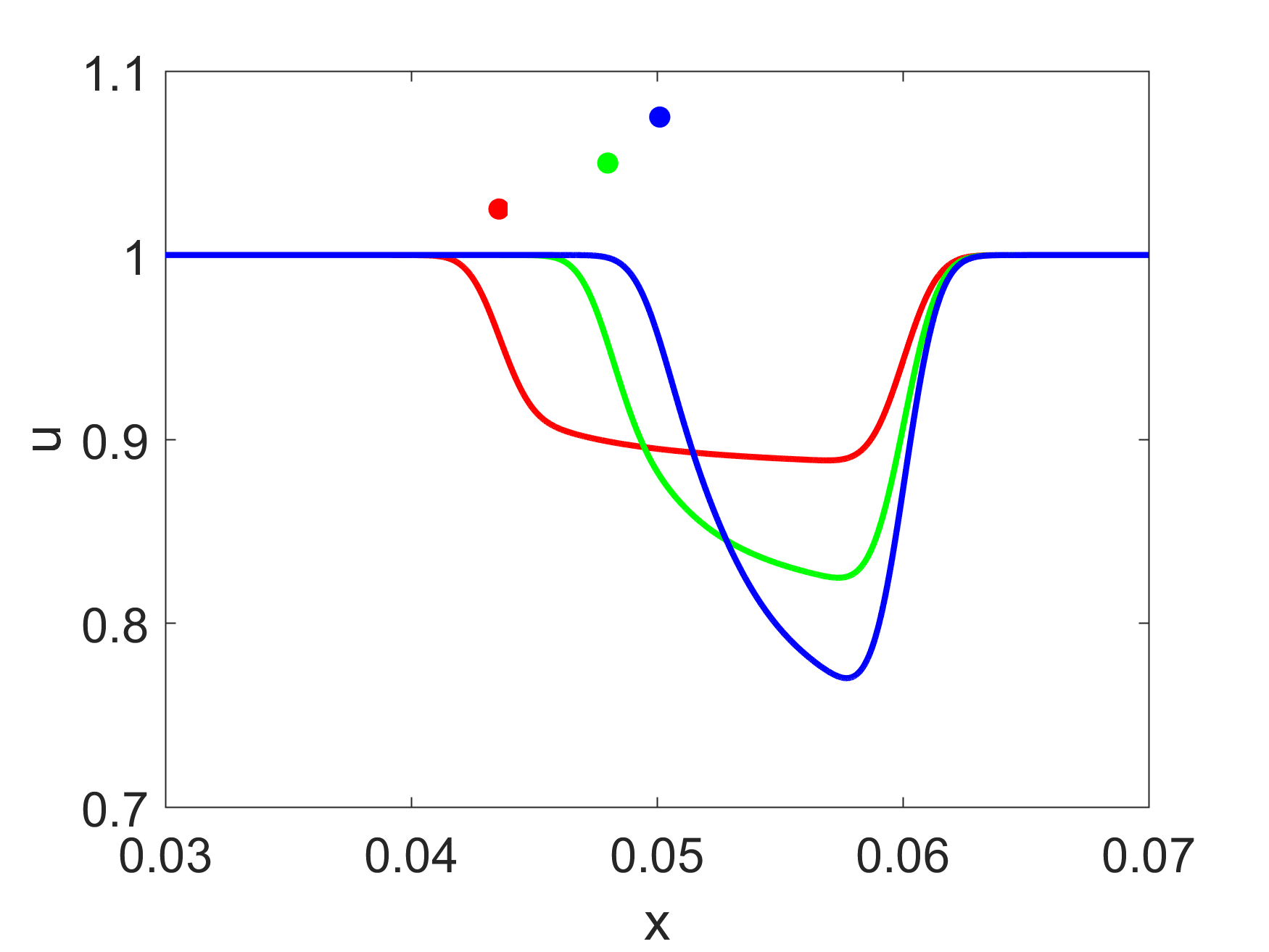}
}
\mbox{
\makebox[0.45\textwidth]{(a)}
\makebox[0.45\textwidth]{(b)}
}
\caption{
(a) Two standard deviation uncertainty bounds for the fluid velocity. The computational particle location is depicted by the blue dot. Red region and the black lines correspond to the method of moment results and the green dashed line to corresponds to MC results. (b) The fluid velocity (solid lines) for three different values of the correction factor ($f_1 = 0.6 (red),1(green),1.4(blue)$) and their corresponding particle position (dots).    
}
\label{fig:ufluid_exp}
\end{center}
\end{figure}

\begin{figure}[htbp]
\begin{center} 
\mbox{
\includegraphics[width=0.45\textwidth]{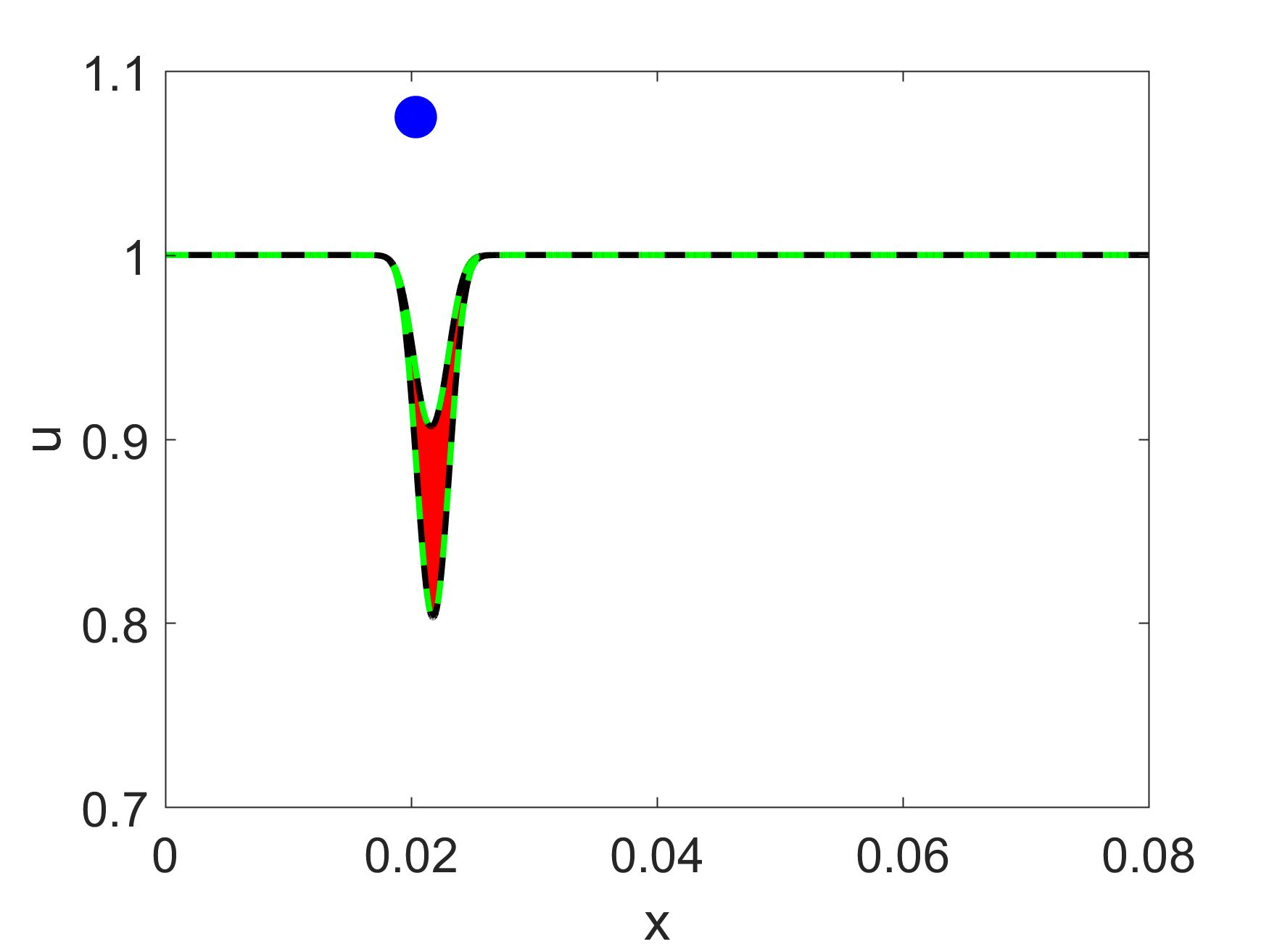}
\includegraphics[width=0.45\textwidth]{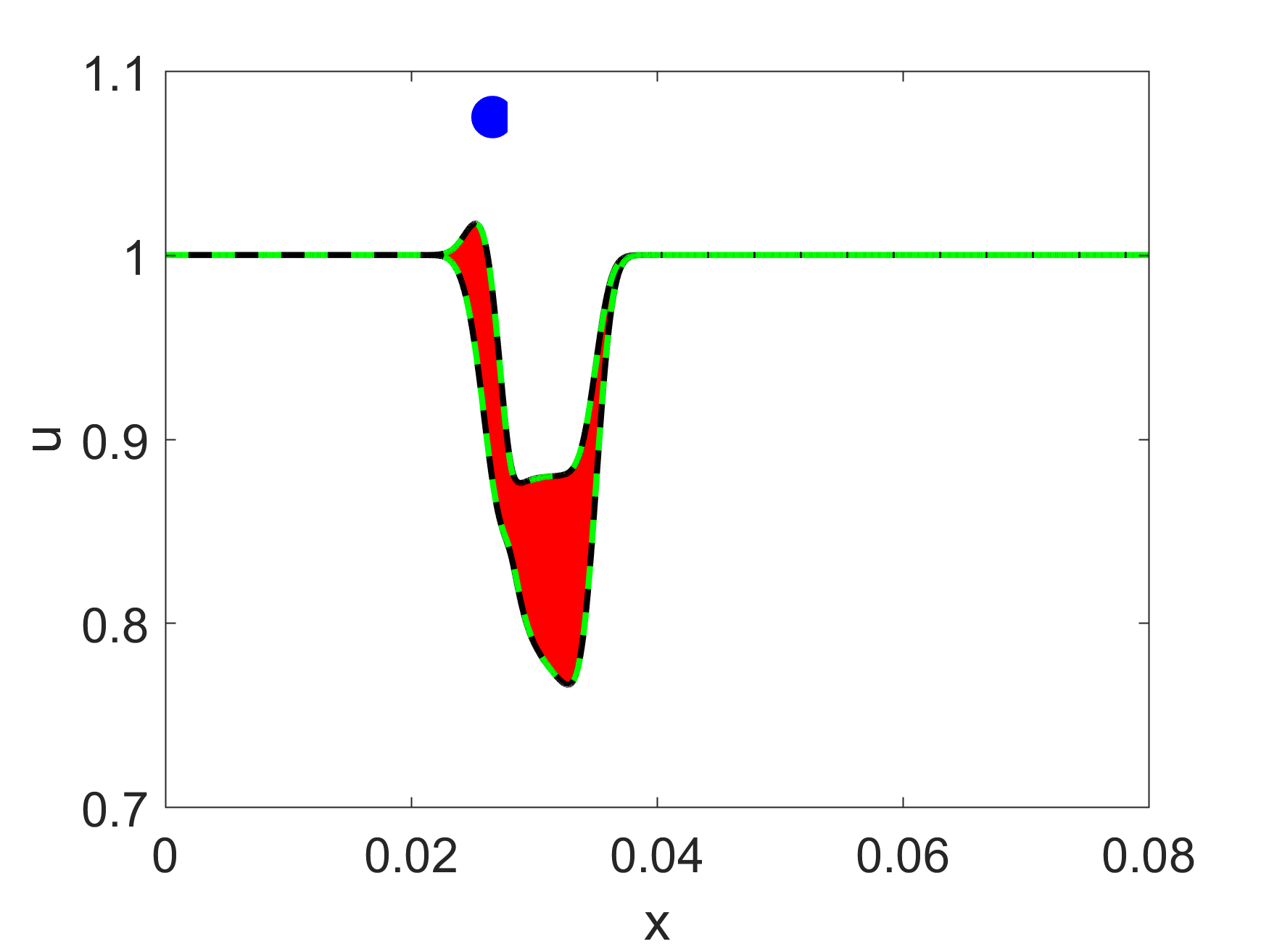}
}
\mbox{
\makebox[0.45\textwidth]{(a)}
\makebox[0.45\textwidth]{(b)}
}
\caption{
Two standard deviation uncertainty bounds for the fluid velocity. The computational particle location is depicted by the blue dot. Red region and the black lines correspond to the method of moment results and the green dashed line to corresponds to MC results.    
}
\label{fig:ufluid_exp_time}
\end{center}
\end{figure}

\begin{figure}[htbp]
\begin{center} 
\mbox{
\includegraphics[width=0.45\textwidth]{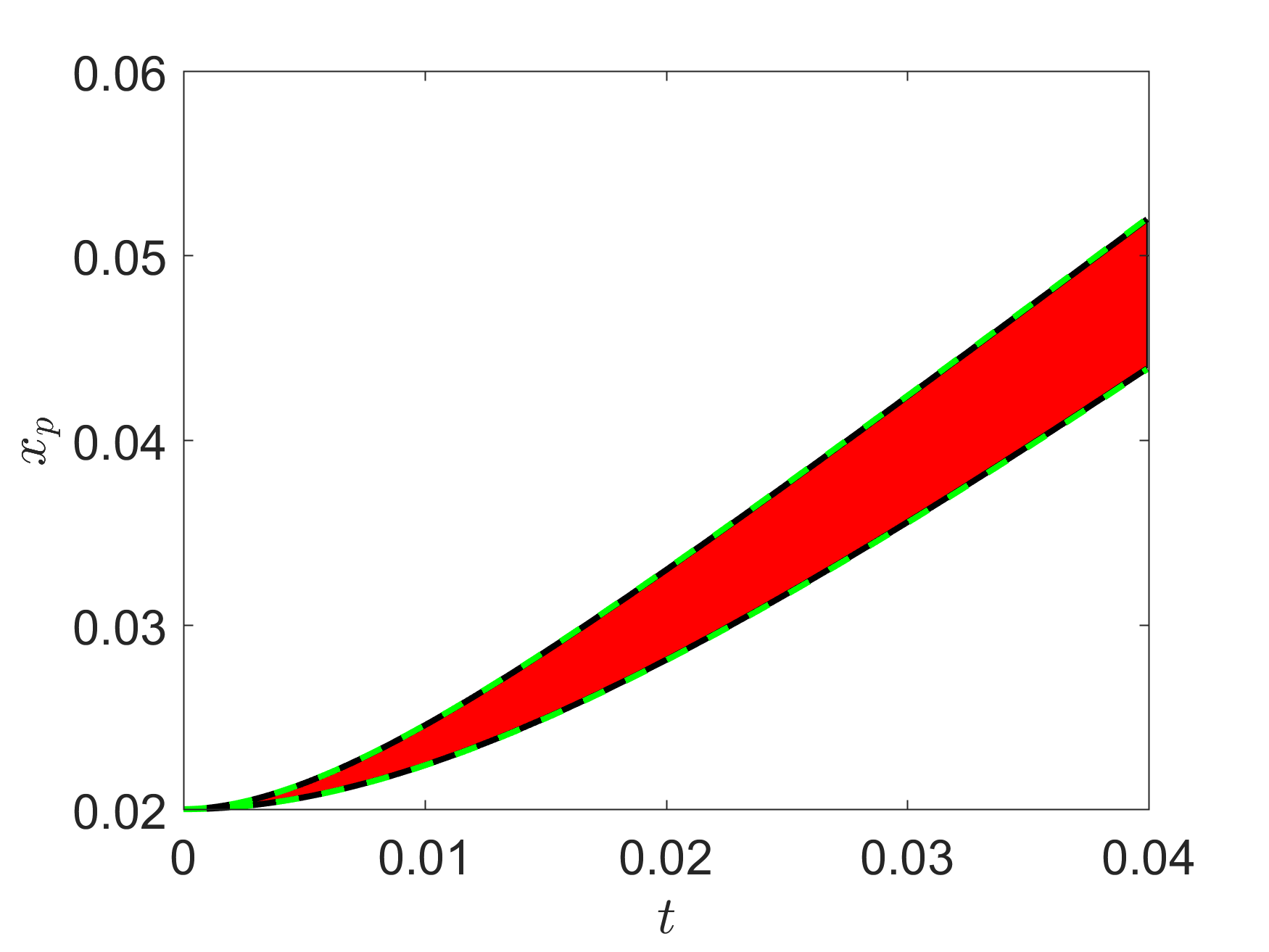}
\includegraphics[width=0.45\textwidth]{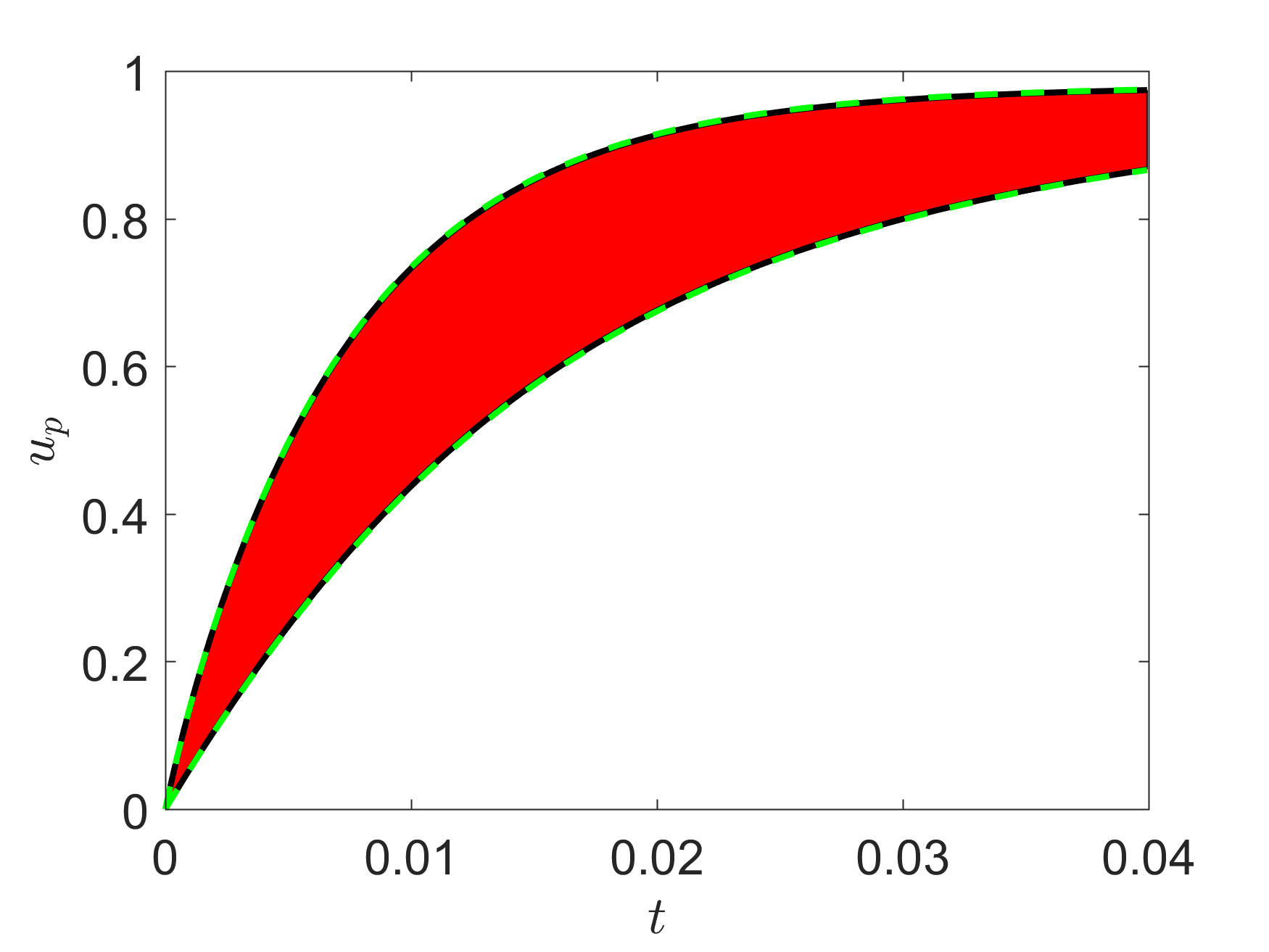}
}
\mbox{
\makebox[0.45\textwidth]{(a)}
\makebox[0.45\textwidth]{(b)}
}
\caption{Red region and the black lines correspond to the method of moment results and the green dashed line to corresponds to MC results. (a) Two standard deviation uncertainty bounds for the particle position. (b) Two standard deviation uncertainty bounds for the particle velocity.
}
\label{fig:xpup_exp}
\end{center}
\end{figure}

We take the MC solution to be the ground truth and compare the results of
moment equations with them. We are able to predict accurate uncertainty bounds
(mean value plus and minus two standard deviations) as shown in Figure
\ref{fig:ufluid_exp} through Figure \ref{fig:xpup_exp}. 

Initially, at the particle location the velocity difference between the particle and
fluid causes the source term to be negative. This results
in sudden reduction in fluid velocity ($u$) at the
initial particle position. This jump in fluid velocity ($u$)  is advected
away from the particle location at the advection velocity
(in our case unity), because the inertia of the particle
yields a lower particle velocity than this
advection velocity.
At later times the particle velocity catches up.
A second jump in fluid velocity ($u$)
coincides with the particle location
and the associated inherent sourcing in the advection equation.
As the particle catches up with the advection velocity, the sourcing
and fluid velocity ($u$) jump reduces.  Over time this results
in the formation of a well like the one shown in Figure \ref{fig:ufluid_exp}b.

Changes in the magnitude of the  correction factor in
\cref{eq:correction} effectively change 
the response time of the particle. This
subsequently controls the well width and particle location as shown in Figure
\ref{fig:ufluid_exp}b. Initially, the uncertainty is localized at the particle
location (Figure \ref{fig:ufluid_exp_time}a)  and is advected over time
according to equation \cref{stochastic_linear}b. At later times, the 
uncertainty has two maxima: the boundaries of the well in fluid velocity ($u$) 
There, the solutions for different correction factors are
significantly different as follows from Figure \ref{fig:ufluid_exp}b  and therefore cause a high
uncertainty. In the area between the boundaries of the well, we notice in Figure \ref{fig:ufluid_exp}b
that the realizations of the solutions are closer to each other, which results in
a region of relatively lower uncertainty.

The moment equations accurately predict the uncertainty of the particle
quantities (Figure \ref{fig:xpup_exp}). The particle position uncertainty keeps
increasing with time. The particle velocity uncertainty increases with time at
the initial stages of the simulation. Eventually, the particle velocity matches
fluid velocity ($u$) so the uncertainty decreases at later times. With shorter
response times (i.e. a larger correction factor), the time interval over which
the particle velocity asymptotically matches the fluid velocity ($u$) reduces, which explains
the trends of the upper and lower bound and the  uncertainty in Figure \ref{fig:xpup_exp}b. 

\begin{figure}[htbp]
\begin{center} 
\mbox{
\includegraphics[width=0.45\textwidth]{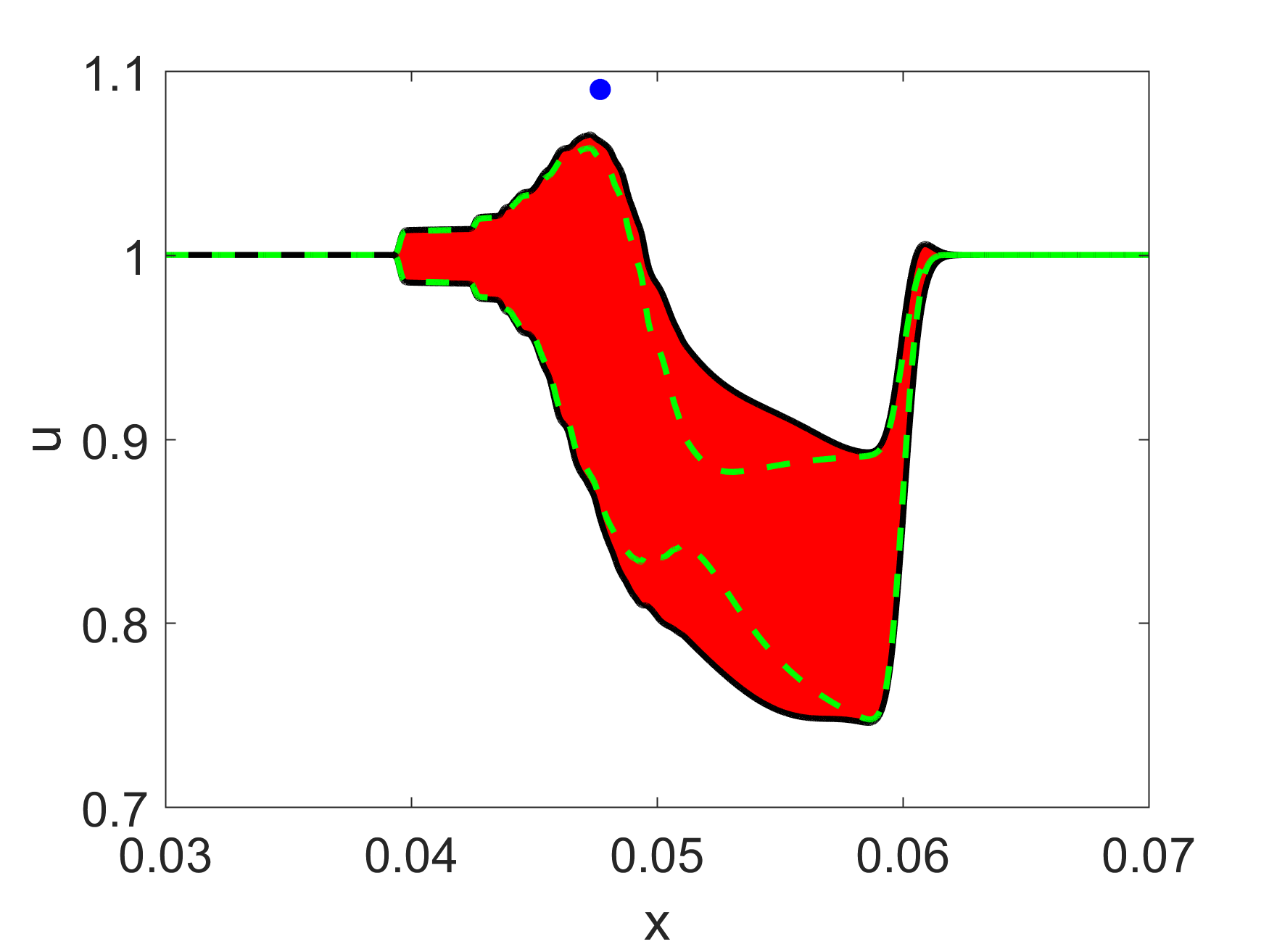}
\includegraphics[width=0.45\textwidth]{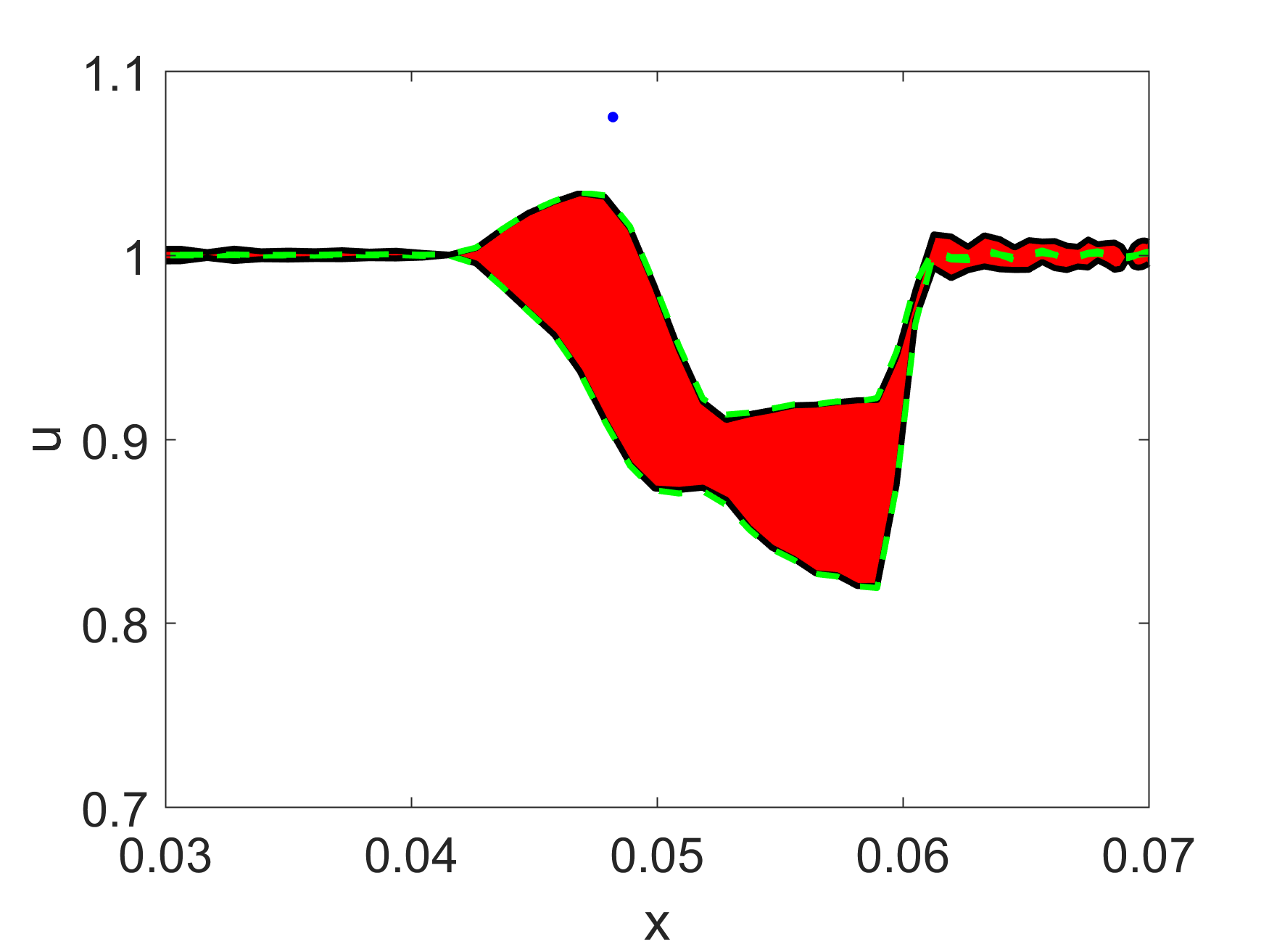}
}
\mbox{
\makebox[0.45\textwidth]{(a)}
\makebox[0.45\textwidth]{(b)}
}
\caption{Red region and the black lines correspond to the method of moment results and the green dashed line to corresponds to MC results. Two standard deviation uncertainty bounds for the fluid velocity. The computational particle location is depicted by the blue dot. (a) using first-order methods (b) using high-order methods
}
\label{fig:exp_10dx}
\end{center}
\end{figure}

If we reduce the support of the exponential source term,
then the first-order method no longer shows
a good match between MC and the method of moments. 
If we take the width
of the Gaussian distribution function to be 10 times the grid spacing,
$\sigma_\delta=10\Delta x$, 
then the mean value, $\overline{u}$, is predicted accurately. 
However, there is a clear inaccuracy in the prediction of the uncertainty 
in \cref{fig:exp_10dx}a.
In \cref{stochastic_linear} we notice that the right hand side
depends on the source term $(S_m)$ and the variation $(u')$. 
By reducing the support of the exponential distribution function, we create
a more singular source term but the numerical method is able to overcome this difficulty and
capture the expected value accurately. Since we have an inaccuracy in the
estimation of the second moment only, we conclude that the source of this
inaccuracy is the high variation of $(u')$ as explained in Remark
5. For the case with the larger support width, the solution is much
smoother and does not cause the variation to be nearly singular.

First-order methods are very diffusive and not very accurate for such problems
so we solve the equations with high-order method as well. We report the result
obtained in \cref{fig:exp_10dx}b. The high-order method solves the inaccuracy
problem and captures accurately the uncertainty. A slight dispersive error
common for high-order method solution with sharp gradients shows in the
form of oscillations in the far field. This is much less of a concern
than the significant diffusive errors of the upwind solution.

\subsection{Advection equation with multiple particles}

Now we consider the advection equation using multiple particles. We use 1601
particles, which are initialized in $[0.01, 0.0132]$ uniformly at rest. The
mass of the particles is set to be $m_p$=0.002. To be consistent with the
accuracy of the space discretization we use linear interpolation to calculate
 the fluid properties at the grid location. In addition, a first-order
distribution function (tent spline) is employed to distribute the particle influence onto the
two closest grid points. 

\begin{figure}[htbp]
\begin{center} 
\mbox{
\includegraphics[width=0.45\textwidth]{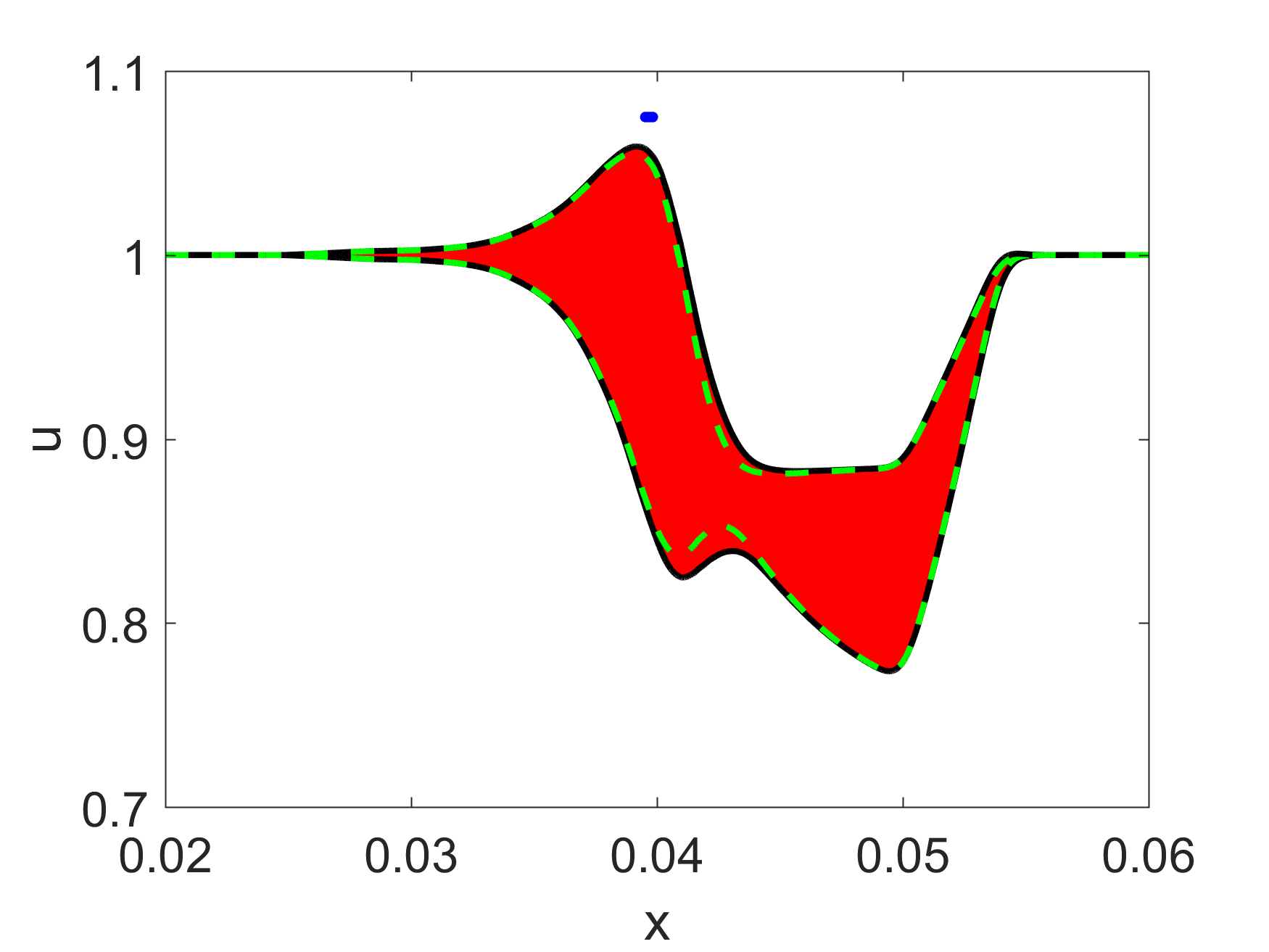}
\includegraphics[width=0.45\textwidth]{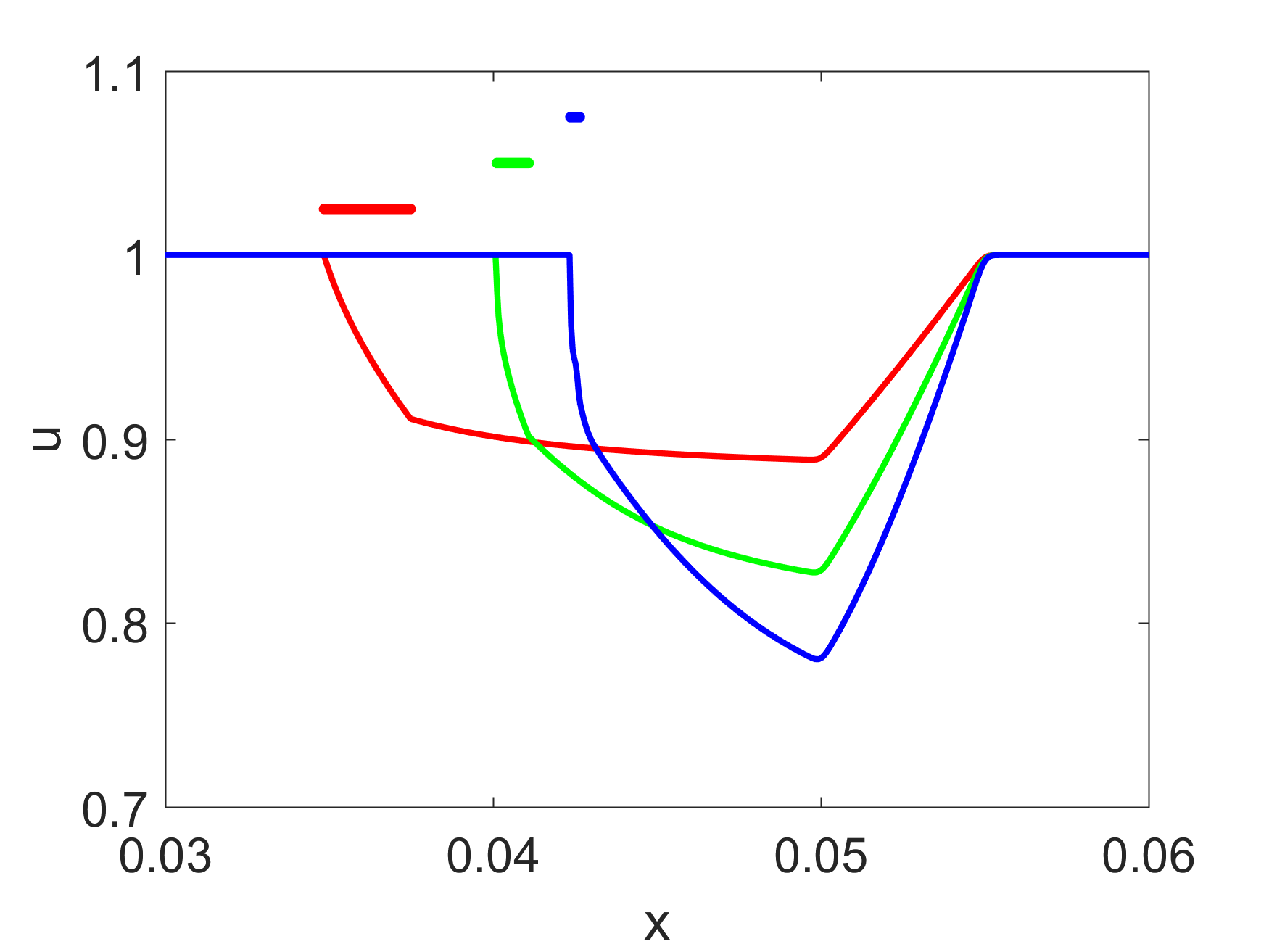}
}
\mbox{
\makebox[0.45\textwidth]{(a)}
\makebox[0.45\textwidth]{(b)}
}
\caption{
(a) Two standard deviation uncertainty bounds for the fluid velocity. The computational particle location is depicted by the blue dot. Red region and the black lines correspond to the method of moment results and the green dashed line to corresponds to MC results. (b) The fluid velocity (solid lines) for three different values of the correction factor ($f_1 = 0.6 (red),1(green),1.4(blue)$) and their corresponding particle position (dots).    
}
\label{fig:ufluid}
\end{center}
\end{figure}

\begin{figure}[htbp]
\begin{center} 
\mbox{
\includegraphics[width=0.45\textwidth]{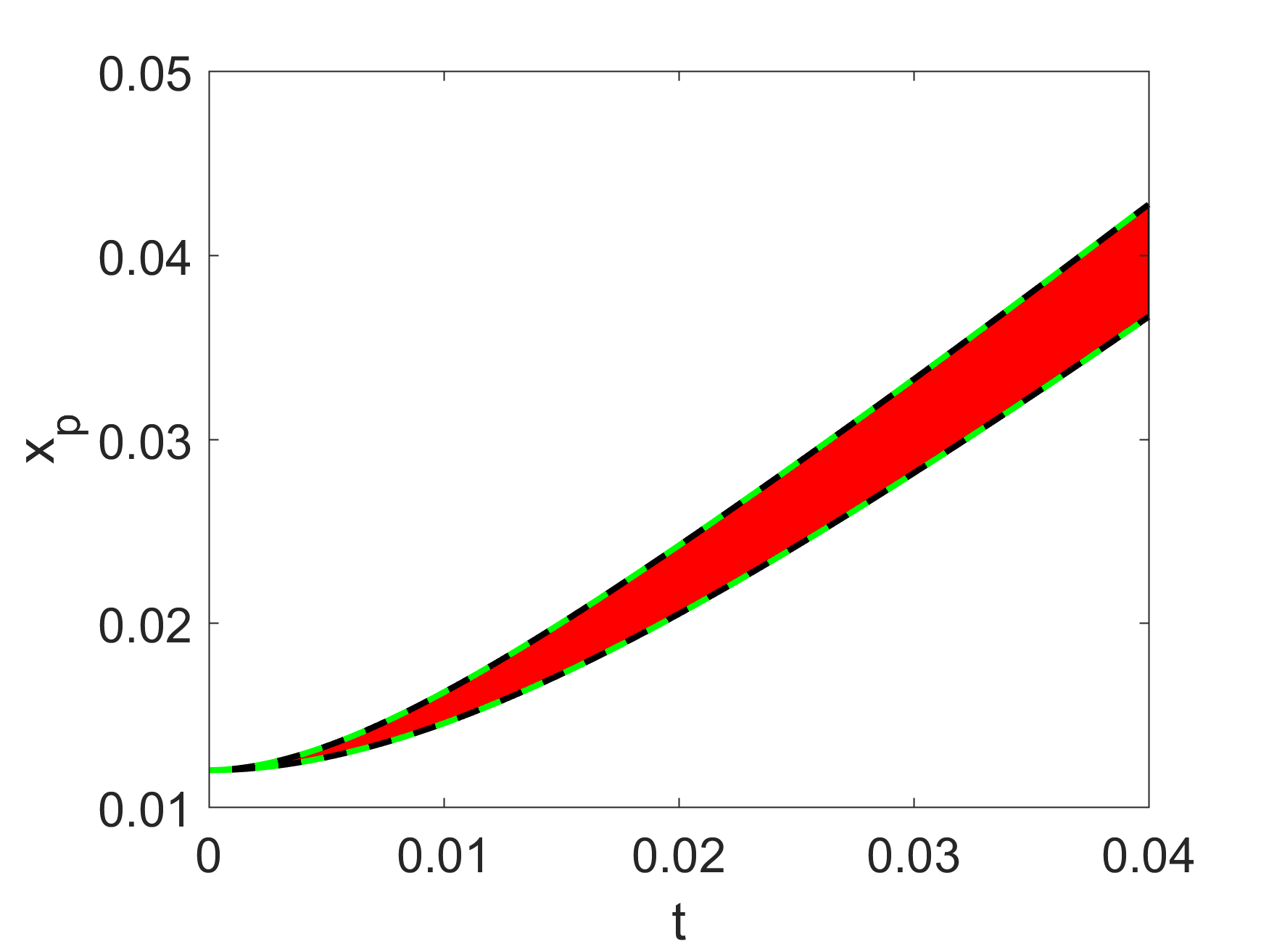}
\includegraphics[width=0.45\textwidth]{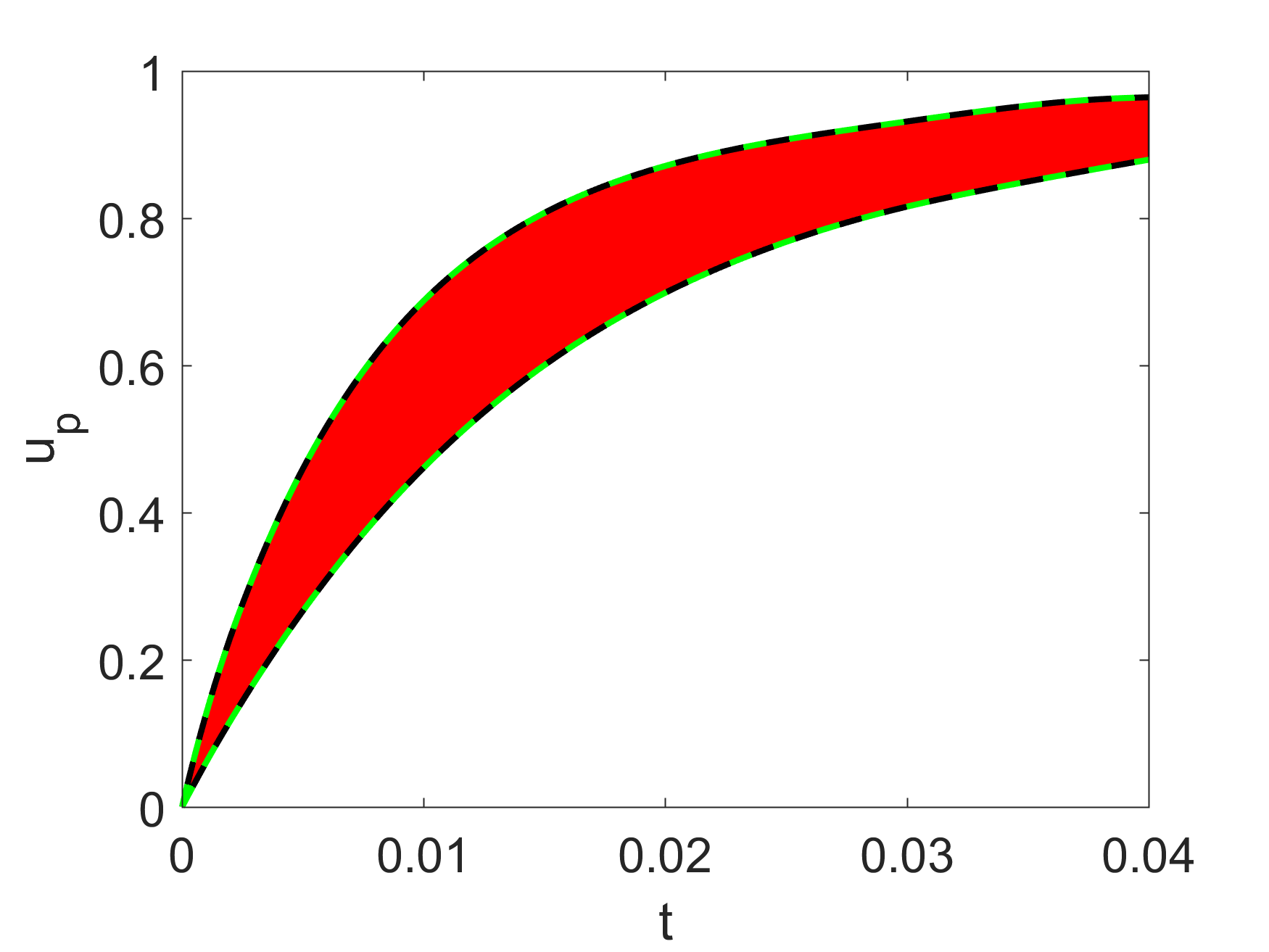}
}
\mbox{
\makebox[0.45\textwidth]{(a)}
\makebox[0.45\textwidth]{(b)}
}
\caption{Red region and the black lines correspond to the method of moment results and the green dashed line to corresponds to MC results. Both plots correspond to a particle in the middle of the cloud. (a) Two standard deviation uncertainty bounds for the particle position. (b) Two standard deviation uncertainty bounds for the particle velocity.
}
\label{fig:xpup}
\end{center}
\end{figure}

The moment equations predict uncertainty bounds that
closely match MC results as shown at Figure
\ref{fig:ufluid} and Figure \ref{fig:xpup}. 
The behavior of the solution and  uncertainty bounds closely
follow the single exponential source case. This is not surprising
since the cumulative effect of the multiple particle is a relatively
smooth sourcing with a wide support.

\begin{figure}[htbp]
\begin{center} 
\mbox{
\includegraphics[width=0.32\textwidth]{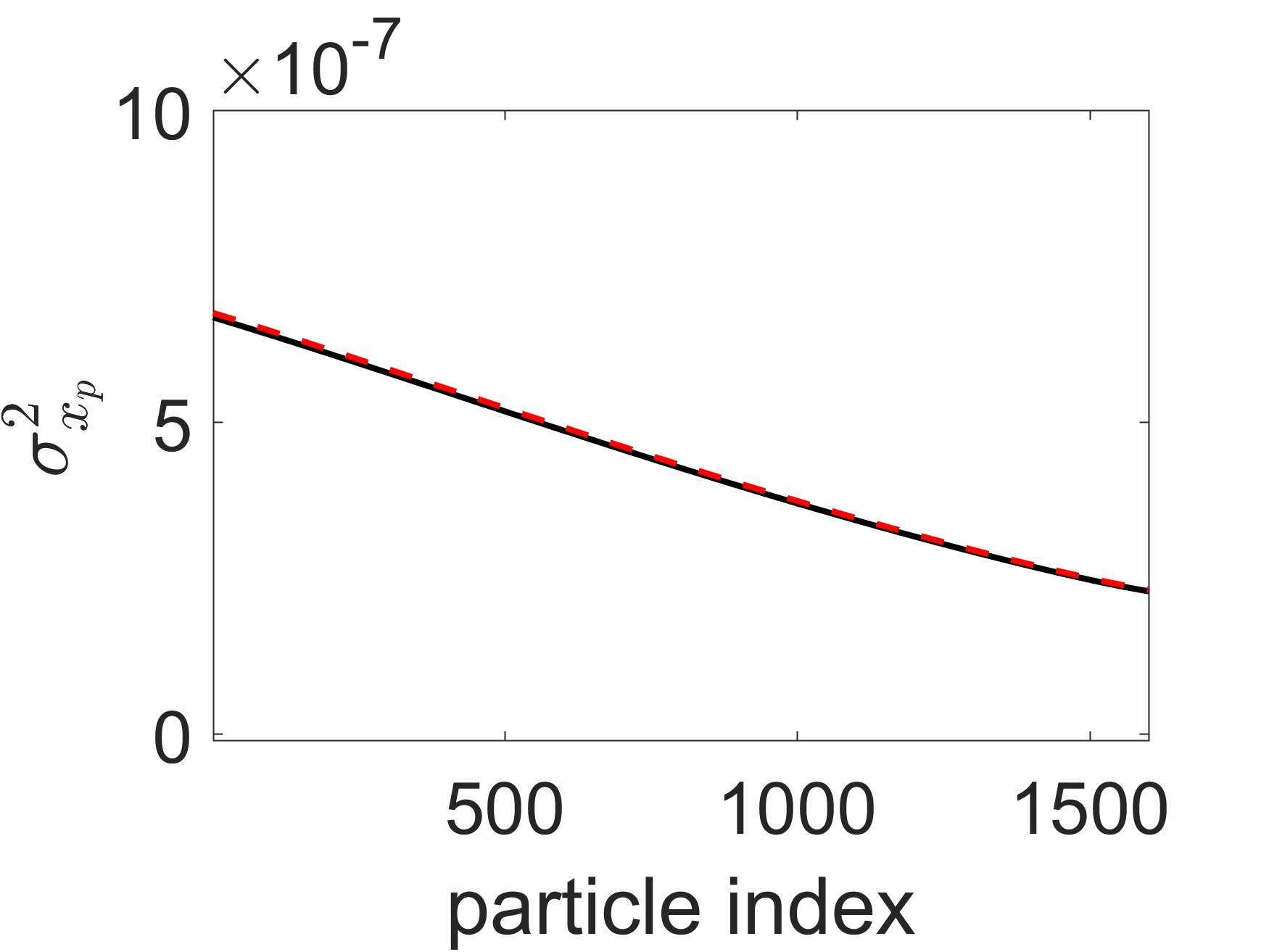}
\includegraphics[width=0.32\textwidth]{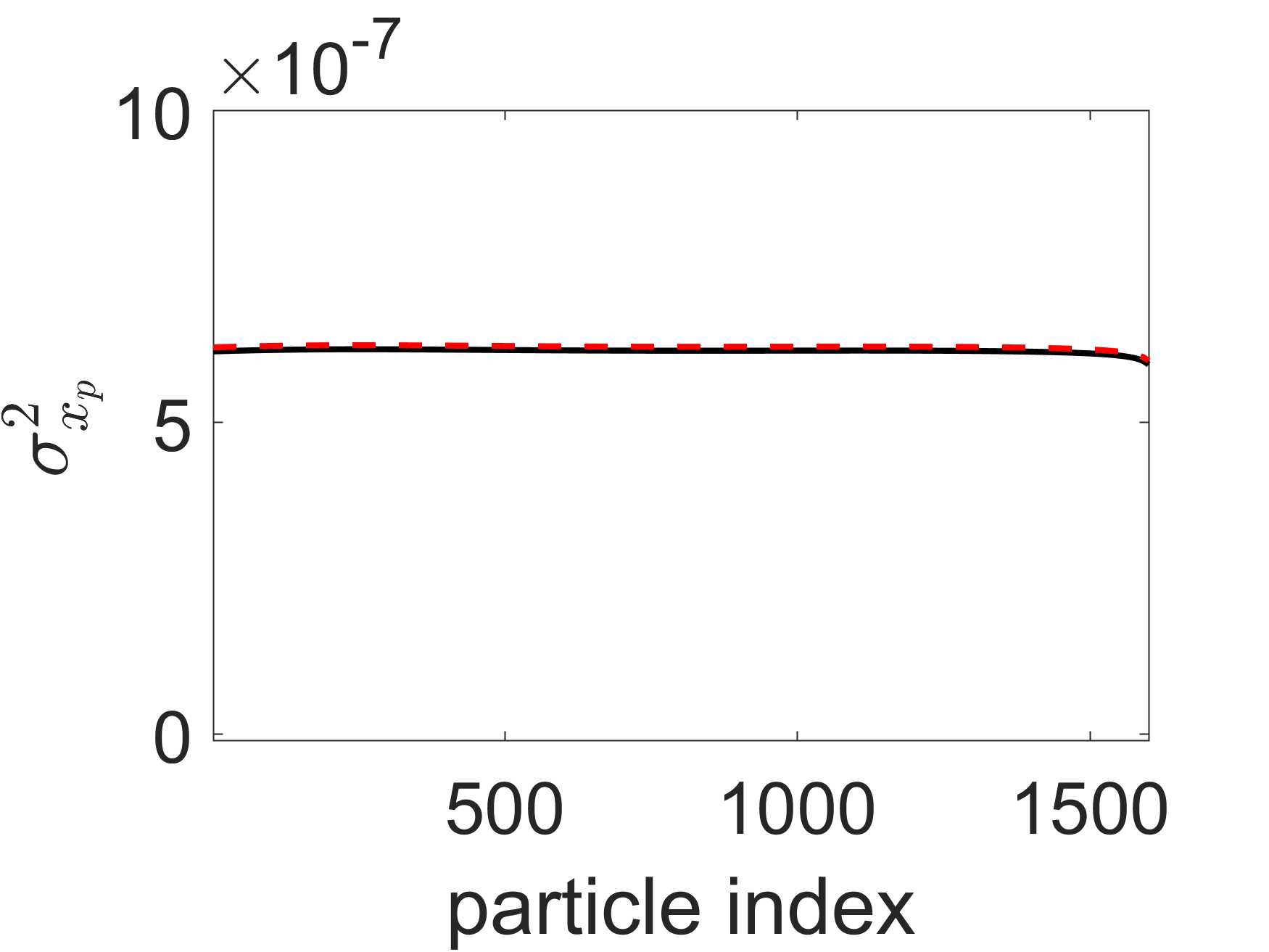}
\includegraphics[width=0.32\textwidth]{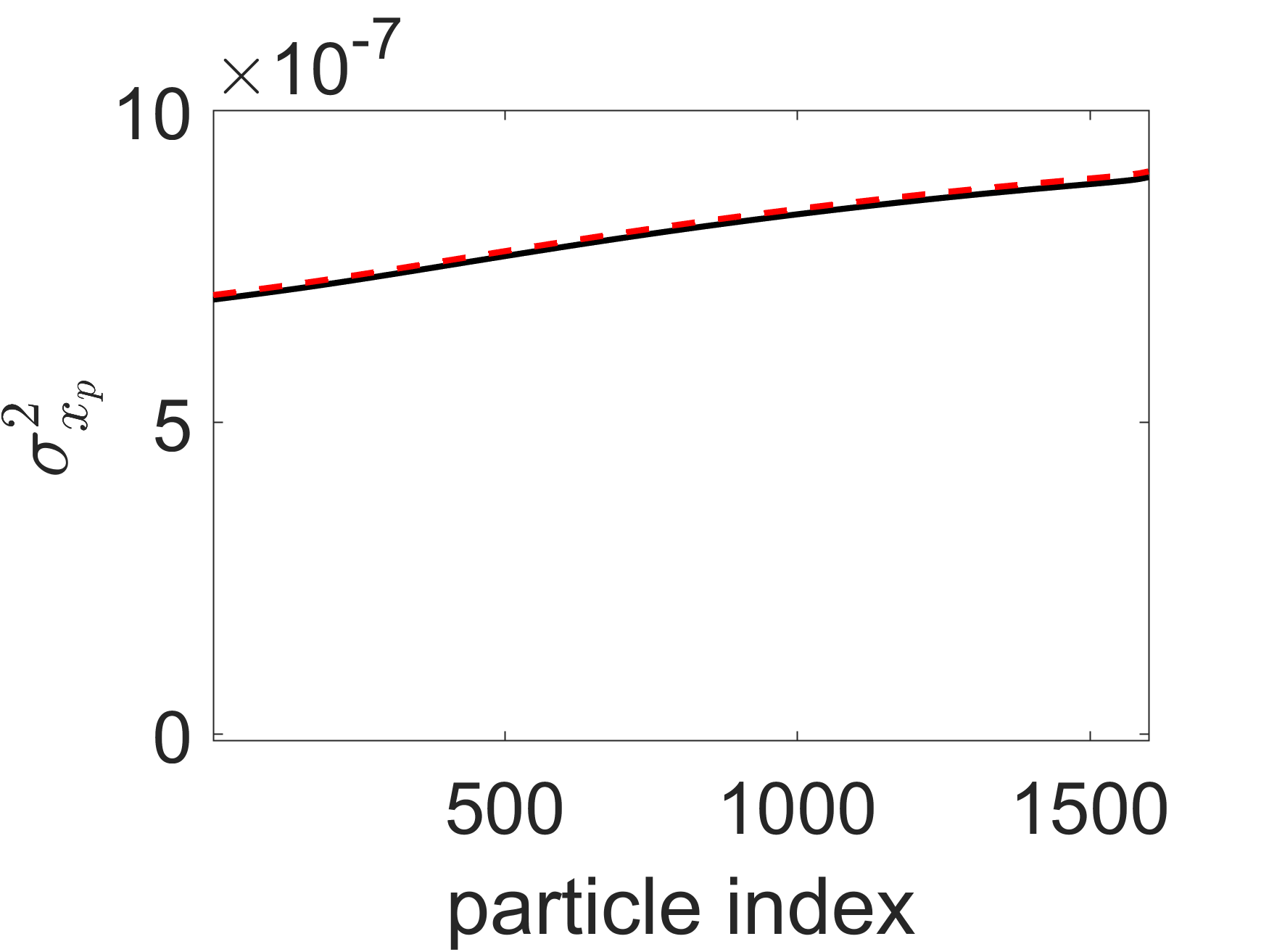}
}
\mbox{
\makebox[0.32\textwidth]{(a)}
\makebox[0.32\textwidth]{(b)}
\makebox[0.32\textwidth]{(c)}
}
\caption{Variance of the particle velocity for the cloud at three different times: a) before trajectory crossing, b) at trajectory crossing, c) after trajectory crossing.
}
\label{fig:up_time}
\end{center}
\end{figure}

One interesting feature of the multiple particles case is that
the particle trajectories cross, i.e.
the particles initially located at the front of the cloud overtake the
particles that are initially located at the rear end of the cloud. This can be
explained as follows. Initially, the front of the cloud faces the 
field, $u$, and reduces its value through its inertial source,
thereby shielding the rear of the cloud. As time advances the particles with
higher velocity keep accelerating more and eventually pass particles that are
at the rear of the cloud. 

The trajectory crossing affect the uncertainty distribution in the particle
cloud. We report the variance of the particle position for each particle in the
cloud in Figure \ref{fig:up_time}. We index them from left to right as they are oriented
in the initial configuration. At initial times the front of the cloud
experiences higher uncertainty because of its greater forcing
till the trajectory crossing occurs. At the
trajectory crossing time, particles are practically at the same location and
the uncertainty is uniformly distributed along the particles. After the
trajectory crossing, the initial rear of the cloud (higher index number) 
has become the front of the cloud. Therefore, it  experiences
 higher incoming fluid velocity ($u$) field resulting in relatively higher
uncertainty for those particles.

\begin{figure}[hbtp]
\begin{center} 
\mbox{
\includegraphics[width=0.45\textwidth]{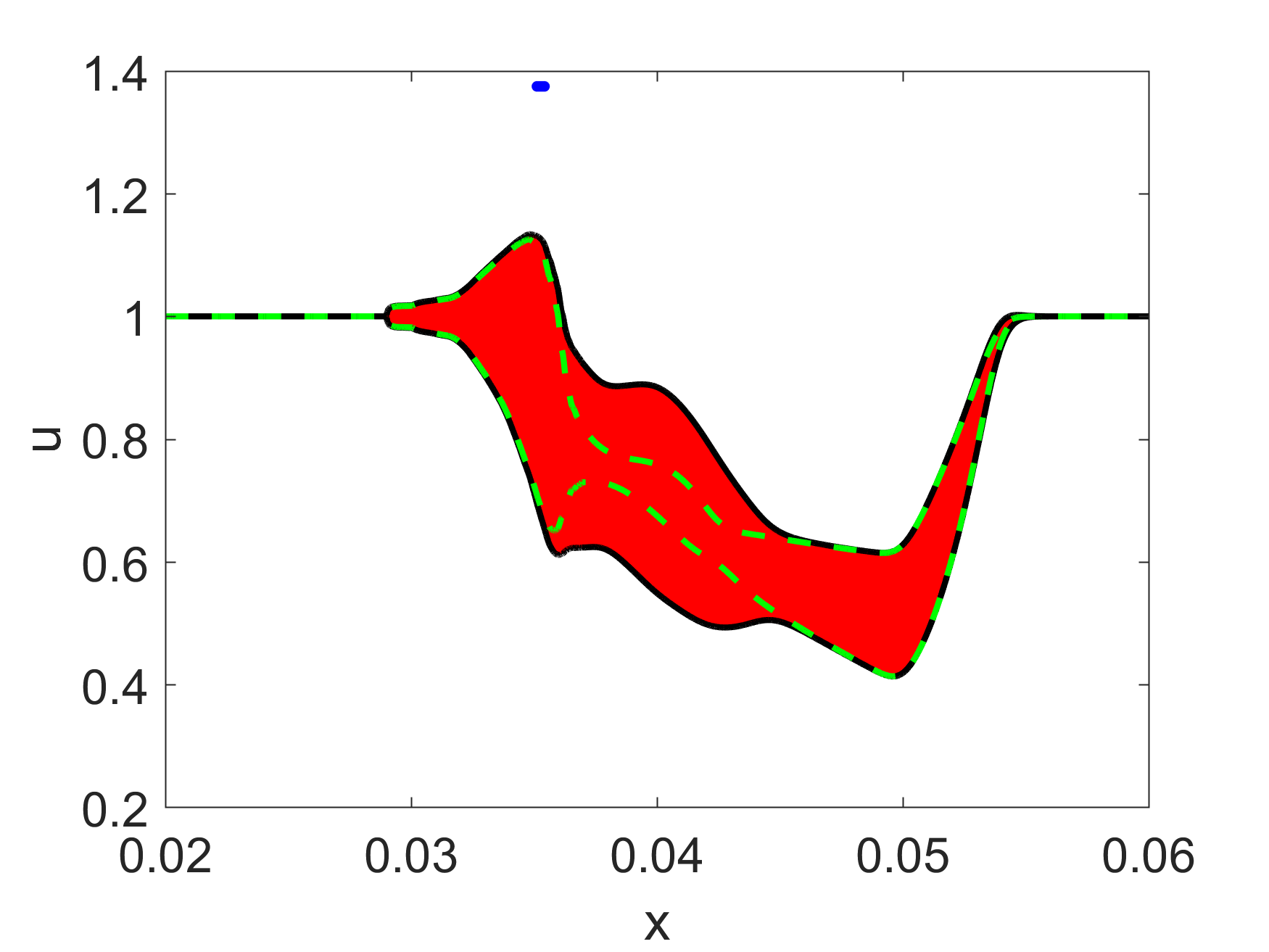}
\includegraphics[width=0.45\textwidth]{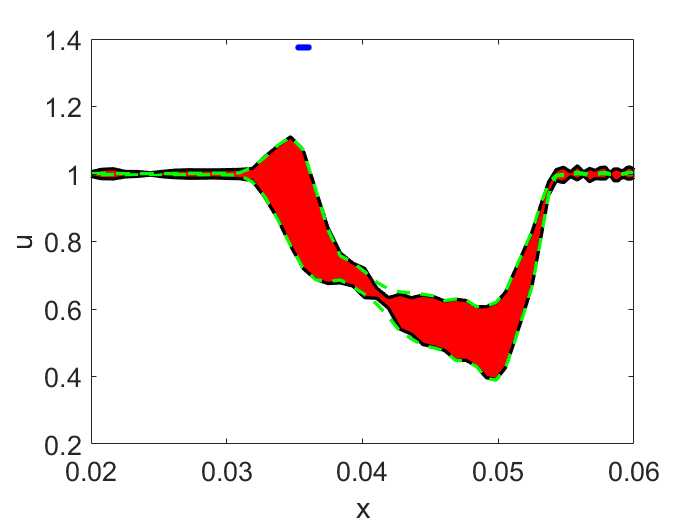}
}
\mbox{
\makebox[0.45\textwidth]{(a)}
\makebox[0.45\textwidth]{(b)}
}
\caption{
Red region and the black lines correspond to the method of moment results and the green dashed line to corresponds to MC results. Two standard deviation uncertainty bounds for the fluid velocity. The computational particle location is depicted by the blue dot. (a) using first-order methods (b) using high-order methods 
}
\label{fig:ufluid1}
\end{center}
\end{figure}

Changing the mass of the particles has a significant effect on the accuracy of
the predicted uncertainty using the method of moments in combination with the first-order
upwind scheme as shown in \cref{fig:ufluid1}a.
Increasing the mass creates a steeper source term which
leads to the same issues we reported on for the single source above.
Again, we are able to capture the
mean value, $\overline{u}$  accurately but not the second moment. 

Using the spectral method results in the same improvements 
over the upwind scheme in the prediction of the uncertainty as shown
in \cref{fig:ufluid1}b. The results show
a more accurate detailed  structure of the uncertainty bounds. Small
oscillations are again observed because of numerical dispersions.

\subsection{Euler equation with multiple particles}

For a more challenging test problem, we consider the Euler equations with multiple
particles. In addition to the singularity challenges that were present in the EL systems of the 
advection equation, the nonlinear Euler equations have inherent shock singularities
even when  the initial conditions are smooth.   This further complicates uncertainty prediction.

Following \cite{jacobs2009high}, 
the particle cloud consists of $4300$ particles that are uniformly distributed
at $[0,0.2981]$. The particle response time and density are $\tau_p =
3.9296\times 10^3$ and a non-dimensional density of $\rho_p  =1200$, respectively. 
We set the Reynolds number to be $Re = 1.7638\times 10^6$. 
We use $600$ grid points to discretize the domain. We assume that
the correction factor follows a Gaussian distribution with a $20\%$ error. The
mean value is determined following \cref{eq:correction}.

\begin{figure}[htbp]
\begin{center} 
\mbox{
\includegraphics[width=0.45\textwidth]{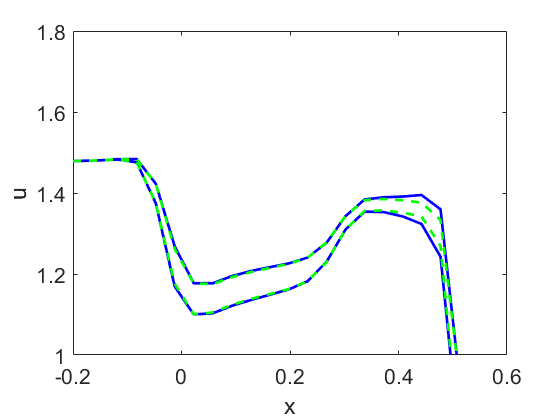}
\includegraphics[width=0.45\textwidth]{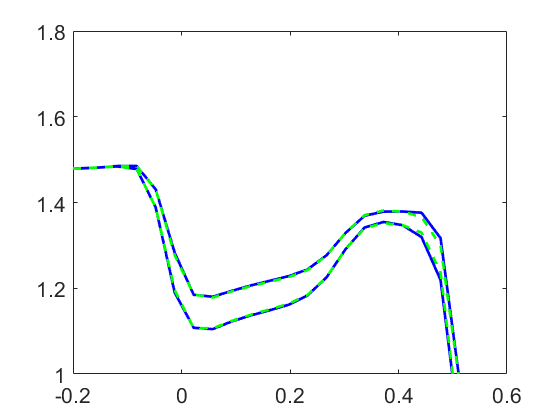}
}
\mbox{
\makebox[0.45\textwidth]{(a)}
\makebox[0.45\textwidth]{(b)}
}
\caption{Two std uncertainty bounds of the fluid velocity at time $t=0.25$ as predicted from the averaged equations (blue line) and MC simulations (dashed green line) for shock Mach number $Ma=2$. (a) using no filtering (b) using filtering
}
\label{fig:euler_uq2}
\end{center}
\end{figure}

\begin{figure}[htbp]
\begin{center} 
\mbox{
\includegraphics[width=0.45\textwidth]{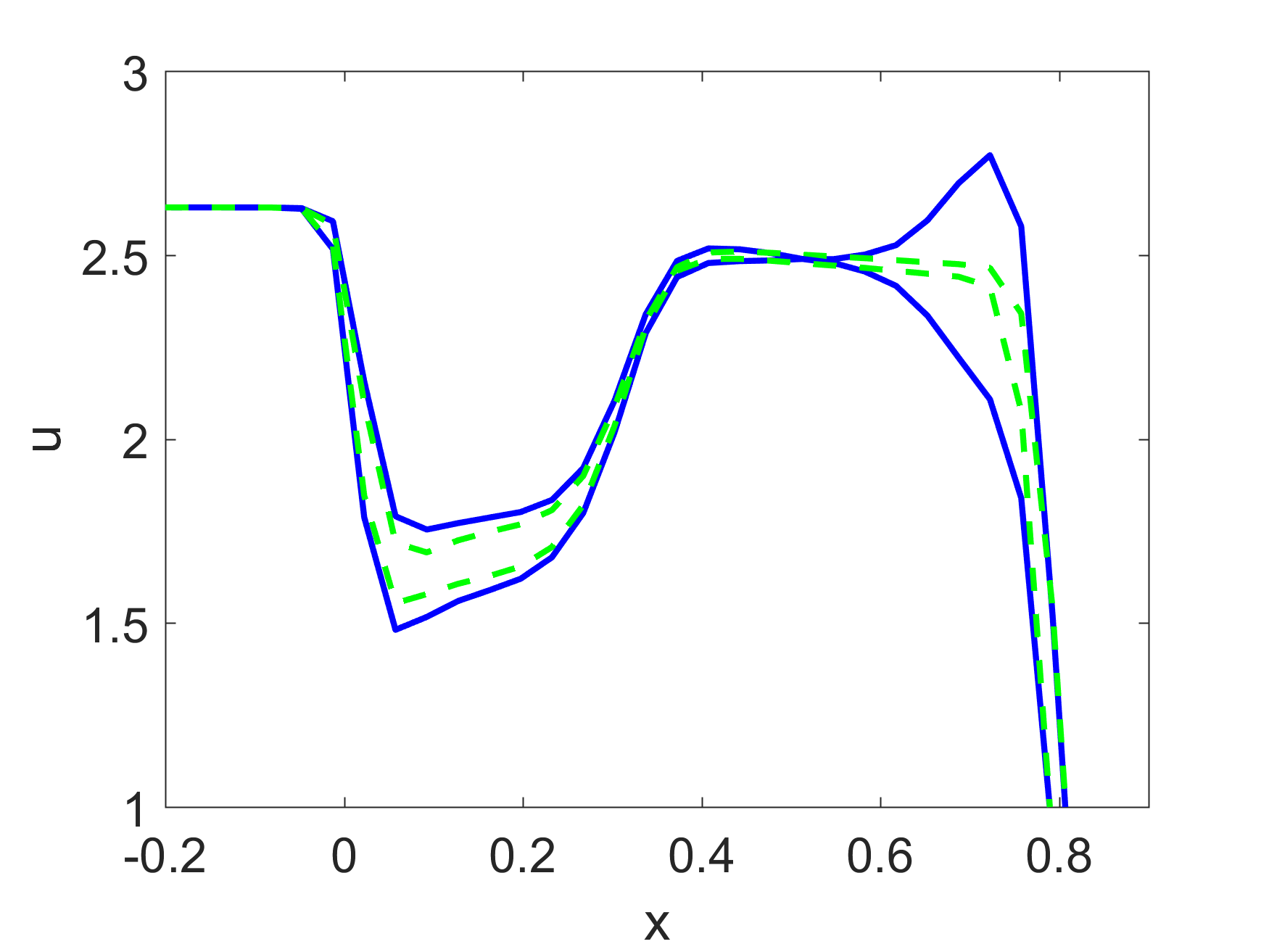}
\includegraphics[width=0.45\textwidth]{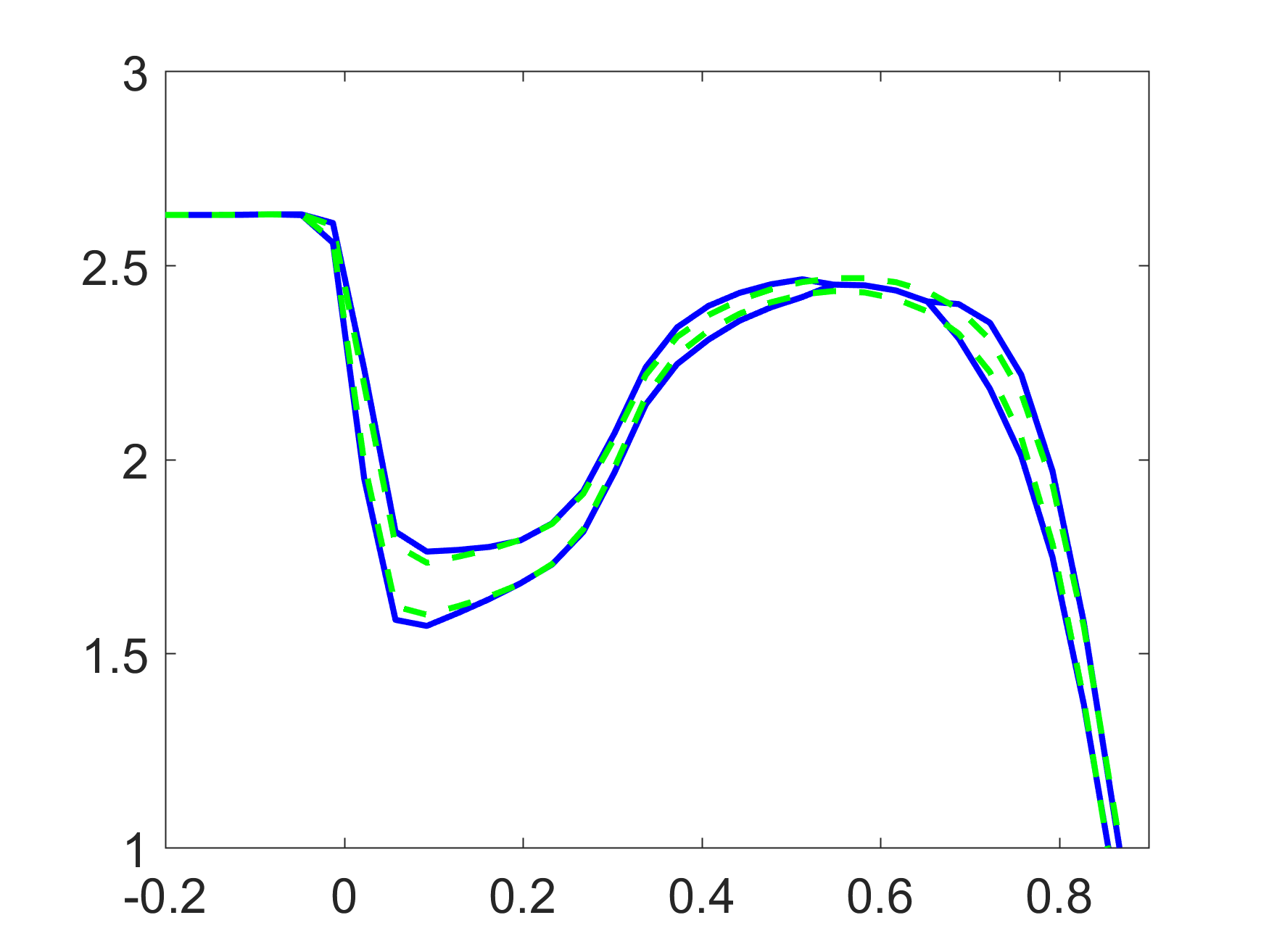}
}
\mbox{
\makebox[0.45\textwidth]{(a)}
\makebox[0.45\textwidth]{(b)}
}
\caption{Two std uncertainty bounds of the fluid velocity at time $t=0.25$ as predicted from the averaged equations (blue line) and MC simulations (dashed green line) for shock Mach number $Ma=3$. (a) using no filtering (b) using filtering
}
\label{fig:euler_uq3}
\end{center}
\end{figure}

We take the MC simulations results to be the ground truth. We report 
results at time, $t$=0.25 for two different shock Mach numbers: $Ma_S=2$ in \cref{fig:euler_uq2}
and $Ma_s=3$ in \cref{fig:euler_uq3}. At this time the shock has passed
through the particles. A reflection shock is also
observed ahead of the cloud. The maximum uncertainty occurs at the location
of the reflected shock and slightly decreases downstream.

We are able to capture accurately the uncertainty bounds using our moment
equations compared to the MC simulations. The only area where there is a mismatch
is at the location of moving shock. We find it  is a localized effect that depends
on the strength of the shock. This is evident if we compare the predictions for
the two different Mach numbers in  \cref{fig:euler_uq2} and
\cref{fig:euler_uq3}. In equation \cref{eq:TKE} we see that most of the right
hand side terms depend on the gradients of quantities. At the location of the
shock these terms become increasingly singular and lead to singular sources
in the equation for $u^{\prime\prime}$. Just like for the advection
case, this renders the moment equation solution inaccurate. 
For this compressible flow case, however, we are already using an accurate fifth-order WENO scheme,
which should prevent the severe dissipation issues noted for the advection equation.
Moreover, switching to the spectral method is not an option since the
inherent shock wave cannot be accurately captured with a spectral method.
Gibbs phenomena not only lead to inaccuracies but also lead to instability of the solution.

Instead, to reduce the impact of the moving shocks
 we choose to  filter the shock and smoothen the solution
 at the shock location to improve our results. Indeed, as shown in \cref{fig:euler_uq2}
and \cref{fig:euler_uq3} this alleviates the large numerical uncertainty
at the shock
front. 

\section{Conclusions}
\label{sec:conclusions}

A method of moments is developed to quantify uncertainty in Eulerian-Lagrangian
models for particle-laden flows. The
method is based on averaging of the deterministic equations
with a stochastic forcing. We 
consider EL models with an 
Eulerian model that is governed by either the advection or Euler equations. 

Comparison of Monte-Carlo simulations with the moment equations closed a priori 
show that the mean and variance are accurately predicted by the
moment model.

If the particle sourcing is large, because of clustering of particles
or because of a relatively high particle mass, strong singular sources
appear in the moment equations. The singularities are caused directly
by the particle sourcing itself, but also
indirectly through strong fluctuations in the stochastic 
Eulerian solution that  in turn appear in  the source terms
of the second moment equation and they have a greater effect. The singular source
terms generate significant uncertainty.

Similarly, (singular) shock waves in the compressible particle-laden flows generate singular sources
in the second moment equation and lead to high uncertainty
in shocked areas.

The singular sources in the moment equations require attention if
solved numerically. In the presence of smooth sources, low-order 
methods are adequate to solve accurately the moment equations. 
However, for singular sources
the low-order schemes are too dissipative to obtain a reasonable match between
Monte-Carlo and the moment equations. A spectral
method improved this match for the linear advection equation.
For the non-linear Euler equation a smoothing through filtering
is necessary to be able to obtain a good comparison between the two approaches.

Our current efforts focus on the closure of the stochastic equations
in a multi-scale framework. This
can be achieved either by creating meta-models using mesoscale simulations in
combination with common closure techniques such as gradient models.

\section*{Acknowledgments}
We gratefully acknowledge the financial support by the Air Force of Scientific Research  under grant number FA9550-16-1-008 and National Science Foundation NSF-CBET Award no.  1603326.


\bibliography{mybibfile}

\end{document}